\documentclass{emulateapj}
\newcommand{\simg}{\gtrsim}

\slugcomment{KEK-TH-1615, KEK-Cosmo-119}
\shorttitle{Opening Angles of Collapsar Jets}
\shortauthors{Mizuta \& Ioka}
\begin{document}
\title{
Opening Angles of Collapsar Jets
}


\author{Akira Mizuta}
\affiliation{
Theory Center, Institute of Particle and Nuclear Studies, KEK, Tsukuba
305-0801, Japan \\
Computational Astrophysics Laboratory, RIKEN, Wako, 351-0198, Japan
}
\author{Kunihito Ioka}
\affiliation{
Theory Center, Institute of Particle and Nuclear Studies, KEK\\
Department of Particles and Nuclear Physics, the Graduate University for
Advanced Studies (Sokendai), Tsukuba 305-0801, Japan 
}

\begin{abstract}

We investigate the jet propagation and breakout from the stellar
progenitor for gamma-ray burst (GRB) collapsars by performing two-dimensional
relativistic hydrodynamic simulations and analytical modeling.
We find that the jet opening angle is given by
$\theta_j \sim 1/5 \Gamma_{0}$
and infer the initial Lorentz factor of the jet at the
central engine, $\Gamma_0$, is a few for existing
observations of $\theta_j$. The jet keeps
the Lorentz factor low
inside the star by converging cylindrically via
collimation shocks under the cocoon pressure, and
accelerates at jet breakout
before the free expansion to a hollow-cone structure.
In this new picture
the GRB duration is determined by
the sound crossing time of the cocoon, after which the opening angle
widens, reducing the apparent luminosity.
Some bursts violating the maximum opening angle
$\theta_{j,\max}\sim 1/5 \sim 12^{\circ}$ imply
the existence of a baryon-rich sheath or a long-acting jet.
We can explain the slopes in
both Amati and Yonetoku spectral relations
using an off-centered photosphere model,
if we make only one assumption that the total jet luminosity is
proportional to the initial Lorentz factor of the jet.
We also
numerically calibrate the pre-breakout model (Bromberg et al.) for
later use.
\end{abstract}

\keywords{(stars:) gamma-ray burst: general ---  hydrodynamics ---
ISM: jets and outflows ---
methods: analytical numerical --- methods: numerical}

\section{Introduction}
\label{sec:intro}
Gamma-ray bursts (GRBs) are the brightest objects in the universe.
The observed isotropic energy
(the apparent energy if it is emitted isotropically)
is of the order of 
or even sometimes more than the solar rest mass energy of
$M_{\odot} c^2 \sim 2\times 10^{54}$~erg.
Current understanding is that the GRB prompt emission is produced by
a relativistic collimated jet,
whose (half) opening angle is $\theta_j \sim 0.1$ 
and Lorentz factor is more than $\Gamma>100$,
significantly alleviating the energy requirements.

The opening angle of a GRB jet is an important quantity
not only for the energetics but also for the event rate of the GRB.
The opening angle also carries information about the central engine.
It is difficult to get any information on the opening angle
from observations of the prompt emission,
which is beamed into an angle $\sim 1/\Gamma$ by a relativistic effect.
The opening angle of a GRB is measured by the light curve
of the afterglow that follows the prompt GRB.
The afterglow light curve exhibits a break
when the jet decelerates to a Lorentz factor of 
$\Gamma \sim 1/\theta_j$, the so-called a jet break,
and we can estimate the opening angle from the jet break time \citep{Sari99}.
Recent observations suggest that the opening angle of long GRBs
are distributed over several to tens of degrees 
(6$^{\circ}$ $\approx$ 0.1 rad; e.g.,~Fong et al. 2012).

What physics determines the opening angle of GRB jets
is not known yet.
In order to understand the physical origin of the jet opening angle,
we have to closely examine 
jet propagation and breakout from the surrounding matter.
Because some GRBs are associated with supernovae
\citep{Galama98,Iwamoto98,Stanek03,Campana06,Mazzali06},
a long GRB is thought to arise from the death of a massive star.
A jet is launched deep inside the progenitor,
and should break out from the stellar envelope to be observed as a GRB.
This is the so-called collapsar model
\citep{Woosley93,MacFadyen99}.
When the jet collides with the stellar envelope,
a shocked jet and a shocked envelope move sideways
from the jet head 
and form a cocoon.\footnote{
In some articles, a cocoon indicates only the shocked jet
that has moved sideway.
However,
the shocked jet is mixed with the shocked stellar envelope
by the shear interaction and loses its identity.
Since it is difficult to define the contact discontinuity,
we use the term ``cocoon'' for both the shocked jet and the
shocked stellar envelope in a broad sense.}
At the expense of the shocked matter,
the jet head moves outward
and finally drills a hole in the stellar envelope.
This is called the jet breakout.\footnote{
To be precise, a jet breakout is different from
a shock breakout that occurs when a forward shock reaches the stellar surface.
This is also different from a jet break seen in the afterglow
light curve.}

There are many simulations of jet propagation
for the collapsar model.
\citet{Aloy00} were the first to show that
a relativistic jet can penetrate the stellar envelope
while maintaining good collimation.
\citet{Zhang03,Zhang04} and \citet{Mizuta06} studied
jet propagation for collapsars
with a wide range of jet parameters,
such as luminosity, initial Lorentz factor, and so on.
\citet{Morsony07} discussed the evolution of
the jet opening angle and indicated that
the opening angle of the jet is relatively smaller than
the initial opening angle.
However the physical origin of the jet opening angle
is still unclear.
What determines the opening angle of the jet?

\begin{figure*}
\epsscale{1.}
\plotone{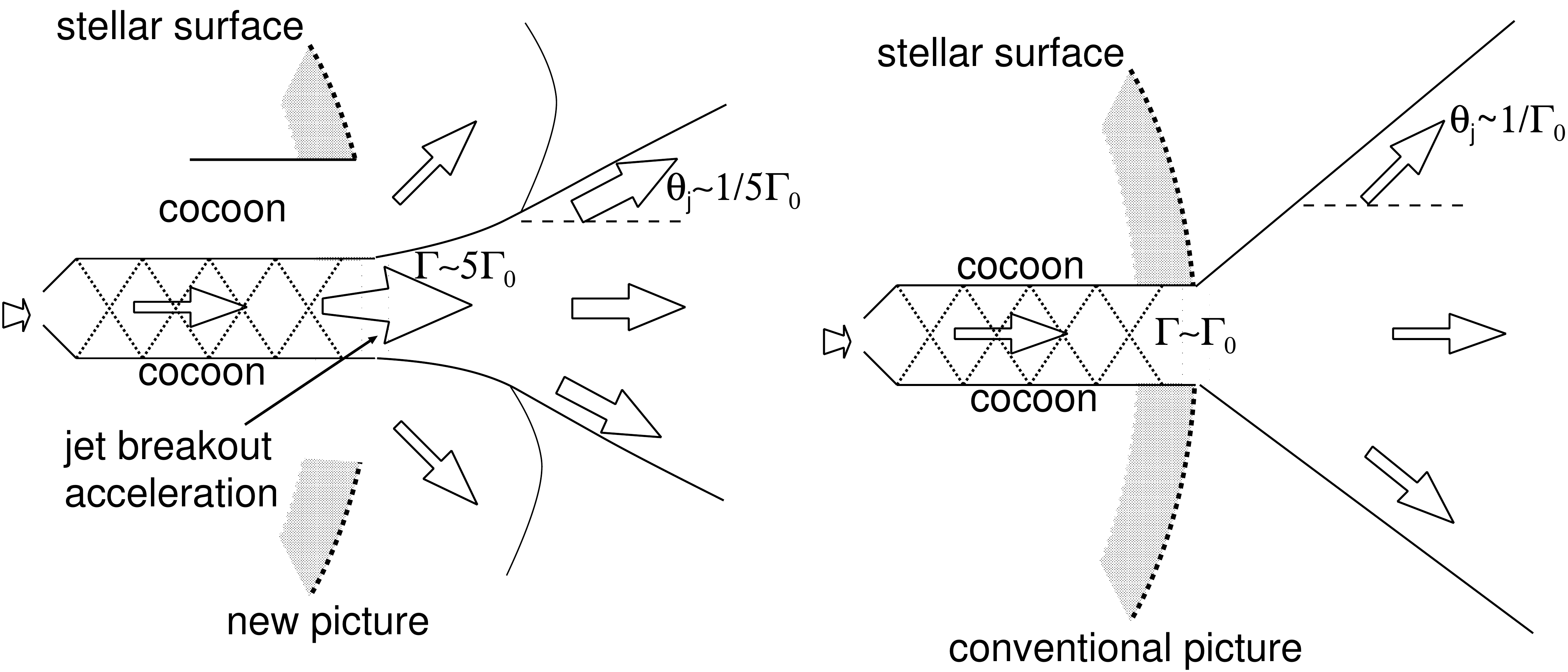}
\caption{\label{fig:break}
Physical picture of the jet
evolution at the jet breakout.
Inside the star, 
the Lorentz factor of the jet keeps the initial value $\sim \Gamma_{0}$
since the jet is almost cylindrical before the jet breakout.
After the jet breakout, the opening angle of the jet is $\sim 1/5\Gamma_{0}$
 (left panel)
rather than $\sim 1/\Gamma_{0}$ (right panel),
because the jet-breakout acceleration occurs at
the jet breakout where
the pressure of the cocoon
decreases outward.
}
\end{figure*}

We consider that the opening angle is given by
\begin{eqnarray}
\theta_j \sim \frac{1}{\Gamma},
\label{eq:thjgab}
\end{eqnarray}
where $\Gamma$ is the Lorentz factor at the jet breakout
from the stellar envelope \citep{Matzner03,Toma07}.
In order to infer $\Gamma$, 
the cocoon pressure is important as it confines the jet
\citep{Matzner03,Toma07,KI+11}.
After passing through the collimation shock \citep{Komissarov97},
the jet becomes cylindrical
since the cocoon pressure is uniform inside the star.
A cylindrical jet is actually observed in simulations 
\citep{Zhang03,Zhang04,Mizuta06,Morsony07,Lazzati09,Lazzati13,Mizuta11}
and recently discussed by \citet{Bromberg11} in an analytical way.
The Lorentz factor of a cylindrical (stationary) jet
is constant ($\Gamma \sim \Gamma_{0}$)
because of flux conservation.
Therefore a naive expectation is that
the opening angle is given by
the inverse of the initial Lorentz factor,
\begin{eqnarray}
\theta_j \sim \frac{1}{\Gamma_{0}},
\label{eq:naive}
\end{eqnarray}
as shown in the right panel of
Figure \ref{fig:break} (the conventional picture).

In this paper, we explore the evolution of the opening angle
for long GRB jet,
and show that the naive expectation
in Equation (\ref{eq:naive}) is partly false.
We perform a series of numerical hydrodynamic simulations
for jet propagation and breakout from collapsars.
We find that the opening angle is a factor $\sim 5$ smaller than
$1/\Gamma_{0}$,
and identify its physical origin 
with the jet-breakout acceleration,
that is, the jet accelerates at the jet breakout
before a free expansion
as shown in the left panel of 
Figure \ref{fig:break} (the new picture).
The jet-breakout acceleration boosts 
the Lorentz factor $\Gamma$
by a factor of $\sim 5$ above than $\Gamma_0$.
Thus Equation (\ref{eq:thjgab}) is correct 
but Equation (\ref{eq:naive}) is not correct.

We also examine the jet dynamics in an analytical way.
We first compare the numerical results with analytical formulae
before the jet breakout by \citet{Bromberg11}.
We calibrate the model parameters with careful numerical calculations,
and make the formulae easier to use.
Then we make an analytical model for the jet-breakout acceleration
after the jet breakout.

In the new picture of Figure \ref{fig:break},
the GRB duration ($T_{90}$)
is determined by
the sound crossing time of the cocoon. After that time, the cocoon
pressure decreases so that the jet is no longer confined by the
cocoon. The opening angle widens from $\theta_j\sim 1/5 \Gamma_0$ (our result)
to $\theta_j\sim 1/\Gamma_0$ (free expansion), reducing the apparent
isotropic luminosity by a factor of $\sim$25. The GRB appears to end even
if the central engine is still active. Therefore the jet-breakout
acceleration is essential for the observed GRB duration.

We infer the initial Lorentz factor of the jet ejected from the central engine
by using the observed opening angles.
We also argue for a possible origin of
the observed spectral relations,
the Amati and Yonetoku relations \citep{Amati02,Yonetoku04}.
We derive the slopes in both the relations
with only one assumption on the jet injection:
that the total jet luminosity is proportional to the initial Lorentz
factor ($L_{j} \propto \Gamma_0$),
under the photosphere model of GRB prompt emission.
These are interesting suggestions
for the emission mechanism of the prompt emission
as well as the jet formation mechanism.

The paper is organized as follows.
In Section ~\ref{sec:method},
we describe
the numerical method and the initial conditions for the jet parameters.
We also introduce the probe particles to measure the jet opening angle.
In Section \ref{sec:results},
we show the main results of the hydrodynamic calculations
and the time evolution of the opening angles.
In Section \ref{sec:analytic},
we present the analytical model of the jet dynamics
and the opening angle of the jet
and compare these quantities with the numerical results.
In Section \ref{sec:discussion},
we apply our results with the observations
of the opening angle, the GRB duration, and the spectral relations
to probe the initial conditions of the jet from the central engine
and the emission mechanism of the prompt emission.
Finally, we summarize our results and give
our conclusions in Section \ref{sec:summary}.

\section{NUMERICAL METHOD}
\label{sec:method}
\subsection{Numerical Scheme}
We have performed two dimensional axis-symmetric relativistic hydrodynamic
simulations of jet propagation before and after the eruption from a
progenitor surface 
in order to obtain the final opening angle of the jet $\theta_j$.
An updated version of
the relativistic hydrodynamic code developed by one of the authors (AM)
is used for the hydrodynamic simulations.
The code solves special relativistic hydrodynamic
equations.
We adopt an adiabatic equation of state, $P=(\gamma -1)\rho \epsilon$,
with a constant specific heat ratio, $\gamma=4/3$,
where $P$ is the pressure,
$\rho$ is the rest mass density, and $\epsilon$ is
the specific internal energy.

The hydrodynamic
code employs the Godunov type fluxes, i.e., an approximate Riemann
solver.
The version of Marquina's flux formula \citep{Donat96}
is used for numerical fluxes.
The second-order accuracy in space is achieved
by the MUSCL method \citep{vanLeer77} and the second-order accuracy in time
is achieved
by the total variation diminishing (TVD) Runge-Kutta method
\citep{Shu88}.
Any radiative processes, such as emission or absorption
of photons are not included in the calculations.
The details of the code and results of  one-dimensional and 
two-dimensional test calculations
are presented
in the appendices of Mizuta et al. (2004,~2006).

\subsection{Grid}
\label{subsec:grid}
Cylindrical coordinate $(z,r)$ are employed
for the hydrodynamic simulations
($z$ is the direction of jet propagation).
In this study, axis-symmetry is assumed for simplicity.
6400 ($z$) $\times$ 500 ($r$) grid points are used.
The high resolution grid points are centered
around the cylindrical axis so that
the interaction with the stellar envelope is accurately captured.
The fine grid points ($10^7~{\rm cm}\times 10^7~{\rm cm}$) are
spaced uniformly at $10^{9}{\rm cm}\le z \le 4\times 10^{10}~{\rm cm}$, 
and $0\le r \le 2\times 10^{9}~{\rm cm}$
which covers the entire jet and a part of
the cocoon at the jet-breakout time.
Logarithmic linear grids
are spaced both in $z$ and $r$ coordinates
at $4\times 10^{10}~{\rm cm} \le z \le 4.3\times 10^{11}~{\rm cm}$
and $2\times 10^{9}~{\rm cm} \le r \le 1.1\times 10^{11}~{\rm cm}$.
At the outer boundaries, the grid sizes are
$\Delta z=7\times 10^8~{\rm cm}$ and
$\Delta r=1.6\times 10^9~{\rm cm}$, respectively.

We perform a resolution study with twice the resolution
of the grids [9728 ($z$) $\times$ 500 ($r$)].
The fine grid points
($5\times 10^6~{\rm cm}\times 5\times 10^6~{\rm cm}$)
are spaced uniformly
at $10^{9}{\rm cm}\le z \le 4\times 10^{10}~{\rm cm}$ 
and $0\le r \le 10^{9}~{\rm cm}$
which covers the jet and some parts of the cocoon at the jet breakout time.
Logarithmic linear grids
are spaced both in $z$ and $r$ coordinates
at $4\times 10^{10}~{\rm cm} \le z \le 6.2\times 10^{11}~{\rm cm}$
and $1\times 10^{9}~{\rm cm} \le r \le 1.1\times 10^{11}~{\rm cm}$.
At the outer boundaries, the grid sizes are
$\Delta z=2\times 10^9~{\rm cm}$ and
$\Delta r=2.2\times 10^9~{\rm cm}$, respectively.
The results of the resolution study are given in Section \ref{sec:resolution}. 

The resolution with ${\Delta z}_{\rm min}={\Delta r}_{\rm min}=10^7{\rm
cm}$ is comparable with
the highest resolution zone in \citet{Morsony07,Morsony10}
who adopted the adaptive mesh refinement technique,
while our high resolution area covers much larger area.
Our resolution study
uses 
${\Delta z}_{\rm min}={\Delta r}_{\rm min}=5\times 10^6{\rm cm}$,
which is one of the highest resolutions so far.

The boundary condition is the free boundary condition
except for the jet injection region and the cylindrical axis.
The reflective boundary condition is imposed at the
cylindrical axis.

\subsection{Stellar Model}
\begin{figure}
\epsscale{1.3}
\rotatebox{0}{\plotone{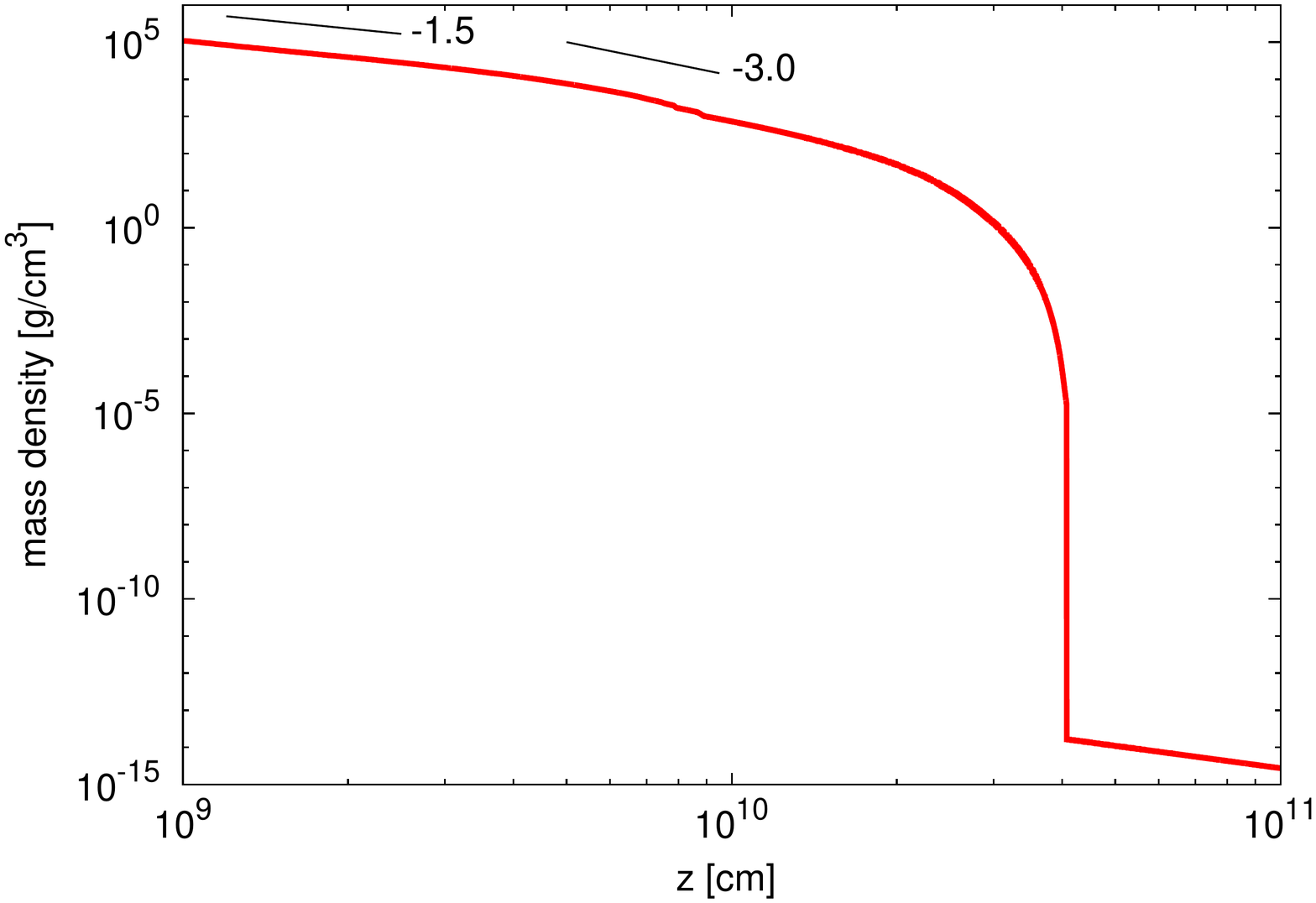}}
\caption{\label{fig:radial_mass}
Initial radial mass density profile (model 16TI in \citet{Woosley06}).
The radial mass density profile is almost power law 
at $10^{9}{\rm cm}\le R \lesssim 6\times 10^{10}~\rm{cm}$ with an index $\sim -1.5$.
Then it quickly drops at $R > 6\times 10^9$~cm.
We extend the gas to the outside of the progenitor
($R> 4\times 10^{10}$~cm) which is
assumed to be very dilute
and a stellar wind profile with the a power-law index $-2$.
}
\end{figure}

We adopt one of stellar models in 
\citet{Woosley06} for the progenitor of the GRB.
The model is named 16TI.
The total mass and radius of the progenitor at the pre-supernova stage
are 13.95 solar mass and
$R_{*}=4\times 10^{10}$~cm, respectively,
where $R$ is the radius in spherical coordinates.
The mass density at the
innermost boundary ($z=10^9$~cm) is about $10^5~\rm{g~cm}^{-3}$.
The radial mass density distribution of the progenitor
is almost a power law with an index $\sim -1.5$ at
$10^{9}{\rm ~cm}\le R\lesssim 6\times 10^9$~cm and
quickly drops at $R > 6\times 10^9$~cm.
The mass density at the last grid of the progenitor ($R=R_{*}$)
is $1.7\times 10^{-5}{\rm ~g~cm}^{-3}$.
Outside the progenitor, we put a low density gas
assuming a stellar wind with a power-law index $-2$, i.e.,
$\rho (R)=1.7\times 10^{-14} (R/R_*)^{-2} {\rm ~g~cm}^{-3}$.
See the radial mass density profile in Figure \ref{fig:radial_mass}.

\subsection{Jet Conditions}
Assuming the jet formation deep inside the progenitor,
we start the numerical calculation
at a distance of $z_{\min}=10^9$~cm from the center of the progenitor.
From the innermost computational boundary, we inlet the jet
into the computational domain.

At least four parameters are necessary 
to characterize the initial condition of the jet.
We choose the cylindrical radius ($r_0$),
the luminosity $(L_j)$, 
the Lorentz factor $(\Gamma_{0})$,
and the specific enthalpy ($h_0 \equiv 1+\epsilon_0/c^2 +P_0/\rho_0$)
of the initial jet,
where the subscript ``0'' stands for the
injection parameter.
Since we inlet the jet parallel to the jet axis
with a small radius $(r_0)$,
it is not necessary to assume an initial opening angle of the jet 
($\theta_0$),
which is determined by the relativistic beaming effect
to be $\theta_0 \approx 1/\Gamma_0$.
This allows us to reduce the number of initial parameters.

\begin{enumerate}
 \item 
We adopt the initial cylindrical radius
of the jet as $r_0=8\times 10^7$~cm.
This value should be sufficiently smaller than
Equation (\ref{eq:rj}), 
which is supposed to be the cylindrical radius 
of the jet after the collimation shock
(i.e., the cylindrical radius of the jet balanced with the cocoon pressure).
If the initial cylindrical radius 
was larger than Equation (\ref{eq:rj}),
the jet dynamics would be different from the true ones
because a thick jet sweeps out a large mass.
Even though $r_0$ should be small,
we still need to cover the jet with a sufficient number of grid points.
In our simulations, 
8 and 16 grid points 
cover the jet at the boundary
for the resolutions,
${\Delta z}_{\rm min}={\Delta r}_{\rm min}=10^7$~cm and
${\Delta z}_{\rm min}={\Delta r}_{\rm min}=5\times 10^6$~cm,
respectively.
Our resolution is high enough to resolve internal structures,
such as shocks and vortices,
inside the jet and the cocoon.

\item
We consider a constant luminosity jet
($L_j=5\times 10^{50}~{\rm erg~s}^{-1}$) for simplicity.
Thus the explosion energy is $\sim 10^{52}$ erg
for a few tens of seconds duration the jet injection,
which is comparable with the values inferred from the observations.

\item
 Another two parameters that define the initial jet are 
the specific enthalpy $(h_0)$
and the Lorentz factor of the jet ($\Gamma_{0}$).
We fix the product of two parameters,
i.e., $h_0 \Gamma_{0}$ to be 538 in this paper.
The product $h_0 \Gamma_{0}$ gives the maximum Lorentz factor
achieved by the adiabatic expansion,
since $h\Gamma$ is conserved by
the relativistic Bernoulli principle along a stream line for
a steady state.
A gas with large enthalpy ($h \gg 1$) expands adiabatically
by decreasing $h$ and increasing $\Gamma$,
i.e., the jet accelerates with a fixed $h\Gamma$.
Our assumption ($h_0 \Gamma_{0}=538$)
satisfies $h_0 \Gamma_{0} > 100$, which is required
for avoiding the compactness problem of the GRB.
The recent {\it Fermi} bursts suggest a relatively large $h\Gamma$
\citep{Abdo:2009a,Ackermann:2010,KI10}.

\item
 The initial Lorentz factors, $\Gamma_{0}=2.5~,5,$ and 10, are studied
to determine the dependence on $\Gamma_{0}$
of the jet dynamics and the opening angle after the jet breakout.
These are the models: G2.5 ($\Gamma_{0}=2.5$),
G5.0 ($\Gamma_{0}=5$), and G10 ($\Gamma_{0}=10$).
We also perform hydrodynamic simulations
with higher resolutions.
These are the models: G2.5H ($\Gamma_{0}=2.5$),
G5.0H ($\Gamma_{0}=5$), and G10H ($\Gamma_{0}=10$).
Since we fix $h_0 \Gamma_{0} (=538)$,
the initial enthalpy ($h_0$) is the smallest ($h_0=53.8$)
for the case with $\Gamma_{0}=10$.
Thus all jets are initially thermal-dominated plasma ($h_0\gg 1$).
The initial specific internal energy ($\epsilon_0/c^2$) is 80 for the model
with $\Gamma_{0}=5$.
As shown later, these initial Lorentz factors
are crucial parameters for the final opening angles.
We set the velocity vector of the jet initially parallel to the $z$-axis.
The jet expands with an initial opening angle $\sim 1/\Gamma_{0}$
as long as the injection angle is less than $\sim 1/\Gamma_{0}$.
\end{enumerate}

\begin{table*}
\begin{center}
\caption{Jet Initial Conditions of the Models
\label{models}}
\begin{tabular}{cccccc}
\tableline\tableline
Model  & ${\Gamma_{0}} ^{~\rm a}$ & Luminosity ($L_j$) & Initial Jet Cylindrical &$h_0\Gamma_{0}(\cong\Gamma_\infty)$ &
${\Delta z}_{\rm min}={\Delta r}_{\rm min}~^{\rm b}$ \\
 & & (erg s$^{-1})$ & Radius ($r_0$) (cm) & & (cm) \\ \tableline
G2.5 & 2.5& & &   &   \\
G5.0 & 5.0& $5\times 10^{50}$ & $8\times 10^{7}$~cm & 538& $10^7~^{\rm c}$ \\
G10 & 10 & & & &    \\ \hline
G2.5H & 2.5& & &   &   \\
G5.0H & 5.0& $5\times 10^{50}$ & $8\times 10^{7}$~cm& 538& $5\times 10^6~^{\rm d}$ \\
G10H & 10 & & &     \\ \hline
 \tableline
\end{tabular}
\end{center}
$^{\rm a}$~{Initial Lorentz factor}\\
$^{\rm b}$~{Highest resolution grid size in the computational domain}\\
$^{\rm c}$~{At the region of $10^{9}{\rm cm}\le z \le 4\times 10^{10}~{\rm cm}$ 
and $0\le r \le 2\times 10^{9}~{\rm cm}$} \\
$^{\rm d}$~{At the region of  $10^{9}{\rm cm}\le z \le 4\times 10^{10}~{\rm cm}$ 
and $0\le r \le 1\times 10^{9}~{\rm cm}$} 
\end{table*}

Table \ref{models} summarizes the initial jet conditions of our models.

\subsection{Probe Particles}
\label{sec:probe}
In order to follow the Lagrange motion of the fluid elements,
we introduce probe particles to trace the path of the fluid elements.
It is necessary to follow the Lagrange motion of the fluid elements
to define the opening angle of the jet,
since the jet opening angle depends on time.
Every 0.01 s, 32 particles
are injected into the computational domain with the jet.
In the injection region at $z_{\rm min}=10^9$~cm,
32 particles are uniformly
spaced at $0\le r\le r_0=8\times 10^7~{\rm cm}$.
In every hydrodynamic time step ($\Delta t$),
the particles move with their local velocities
calculated by the hydrodynamic simulation, i.e,
$\mbox{\boldmath $x$}_{\rm new} =
\mbox{\boldmath $x$}_{\rm old}+\mbox{\boldmath $v$}\Delta t$,
where $\mbox{\boldmath $x$}$ is the position of a particle
and $\mbox{\boldmath $v$}$ is the local velocity.

Tracing the Lagrangian motion of the fluid elements
allows us to find the location where 
the free expansion starts after the jet breakout.
Since we define the jet opening angle
by the direction of free expansion (see, Section \ref{sec:opening_angle}),
the Lagrangian motion is
very important for the quantitative analysis,
especially for finding a relationship between
the opening angle and the Lorentz factor.
The particle path is also useful for identifying
whether the fluid elements have moved into the cocoon or not.

\section{RESULTS}
\label{sec:results}
\subsection{Overall Evolution}
\begin{figure}
\epsscale{1.2}
\plotone{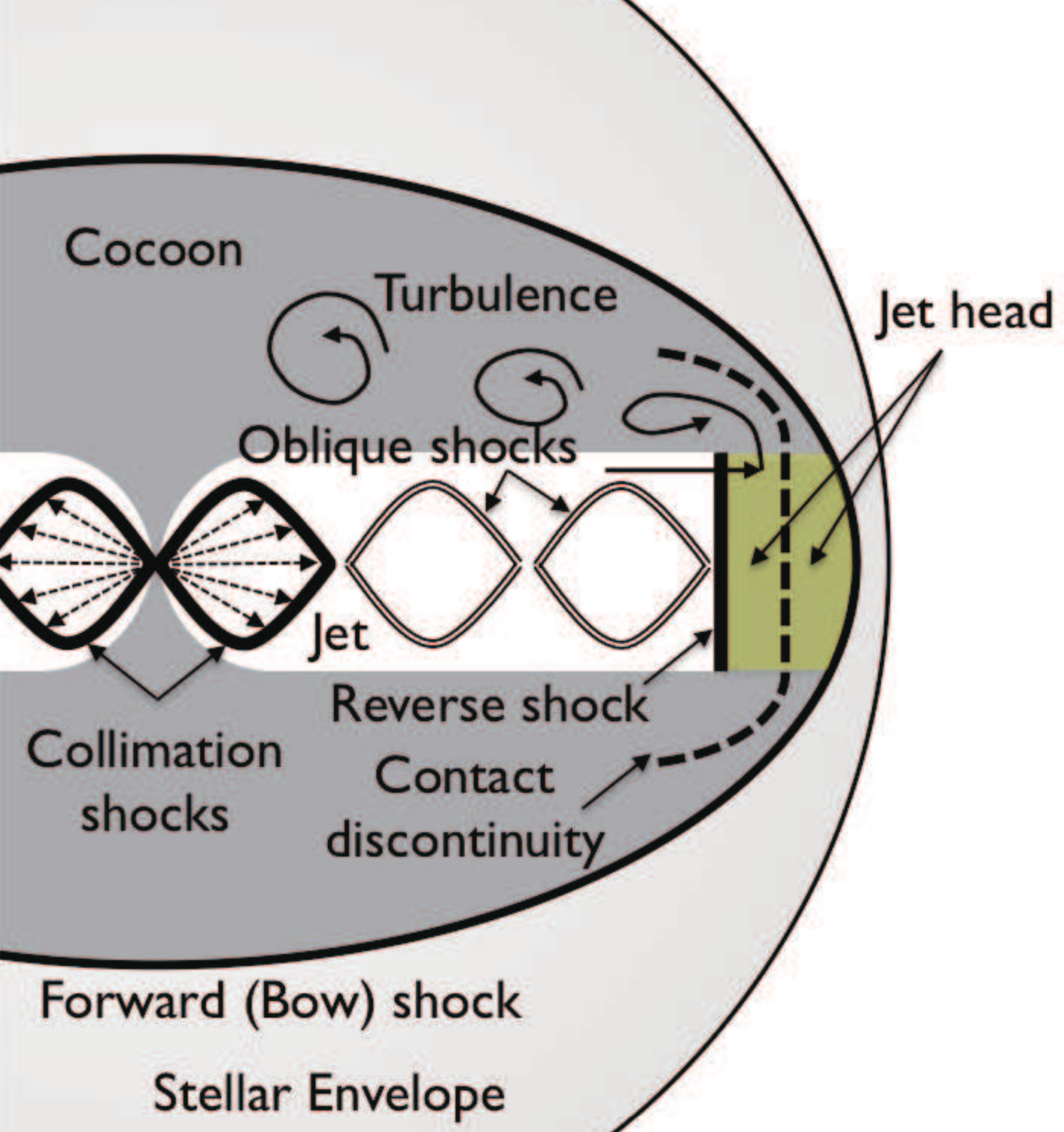}
\caption{\label{fig:physicsmap}
Schematic figure of the jet while the forward shock is inside the progenitor star.
}
\end{figure}

The jets in all models of Table \ref{models}
successfully drill through the stellar envelops.
Before the jet breakout,
the jet hits the reverse shock
that is produced by the interaction between the jet
and the stellar envelope.
Then, the shocked jet moves sideway,
forming a high pressure cocoon,
as shown in the schematic Figure~\ref{fig:physicsmap}
\citep{RamirezRuiz02}.
The high pressure in the cocoon works to confine the jet.
The jet is confined through a collimation shock 
that deflects the velocity vector 
into a direction parallel to the cylindrical axis.
The high pressure cocoon is also discussed 
in the context of Fanaroff-Riley type II jets \citep{Begelman89}.

When the forward shock reaches the stellar surface,
the shocked stellar envelope starts to expand into 
the circumstellar matter, i.e., shock breakout occurs.
Soon the jet also starts to expand
into the circumstellar matter, i.e., jet breakout occurs,
as presented in previous numerical simulations 
\citep{Aloy00,Zhang03,Zhang04,Mizuta06,Morsony07,Mizuta09,Mizuta11,Nagakura11}.
After jet breakout, the jet advances
in the circumstellar matter which is assumed to be very dilute.

\subsection{Before the Jet Breakout}
\label{sec:before}
\begin{figure}
\epsscale{1.7}
\plotone{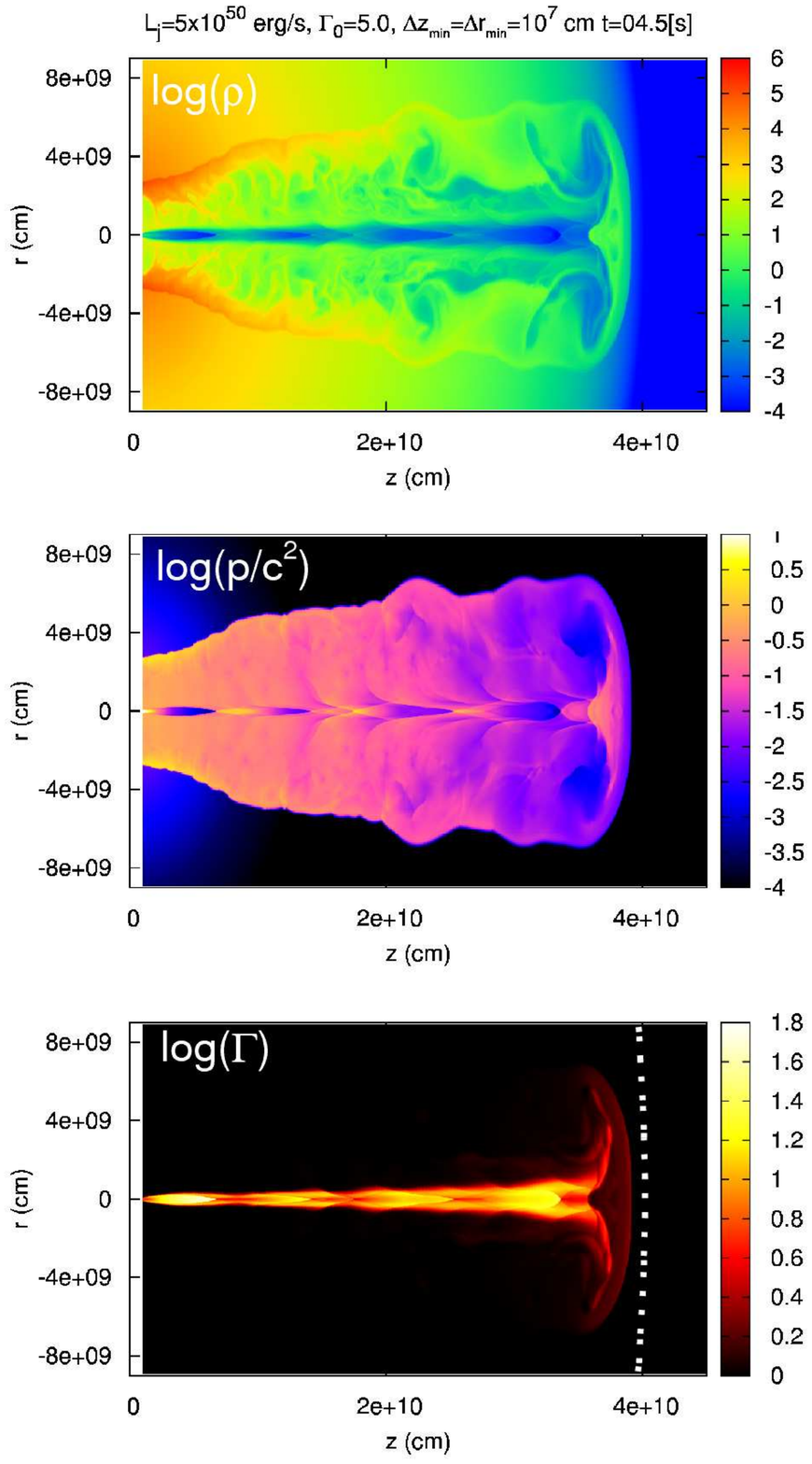}
\caption{\label{fig:Gam0_5_0045}
Mass density in ${\rm g~cm}^{-3}$
(top), pressure in ${\rm dyn~cm}^{-2}/c^2$ (middle),
and Lorentz factor (bottom)
contours of the model
$\Gamma_{0}=5$ at $t=4.5$~s (model G5.0).
Note that the aspect ratio of $z$ and $r$ is not unity
in order to enhance the fine structures in the jet and the cocoon.
The white dashed line in the Lorentz factor contour
indicates the initial stellar surface.
}
\end{figure}

The jet is well confined by the high pressure cocoon
before the jet breakout.
Figure \ref{fig:Gam0_5_0045} shows 
the contours of the mass density, the pressure and the Lorentz factor
of the jet just before the jet breakout
for the model G5.0 ($\Gamma_{0}=5$).
We should note that the $r$ axis is elongated and
the aspect ratio of $z$ and $r$ is not unity
in order to highlight the fine structures in the jet and the cocoon.
Although we inlet a jet with a velocity
parallel to the cylindrical axis, the jet tries to expand
with an opening angle of $\theta_0\sim {\Gamma_{0}}^{-1}$
because the thermal energy is large.
Around the injection point, the Lorentz factor rapidly increases
and then drops across a discontinuity.
Since the pressure and the mass density increase across this discontinuity,
it is the shock surface.
The shock deflects the velocity vector almost towards the jet axis,
and is the so-called collimation shock.
The collimation shock is produced by the interaction between
the expanding jet and the high pressure cocoon
\citep{Komissarov97,Komissarov98,Bromberg09,Bromberg11}.
To distinguish it from the other collimating oblique shocks in the jet,
we sometimes call it the first collimation shock.
See the schematic Figure~\ref{fig:physicsmap}.

\begin{figure}
\epsscale{1.3}
\rotatebox{0}{\plotone{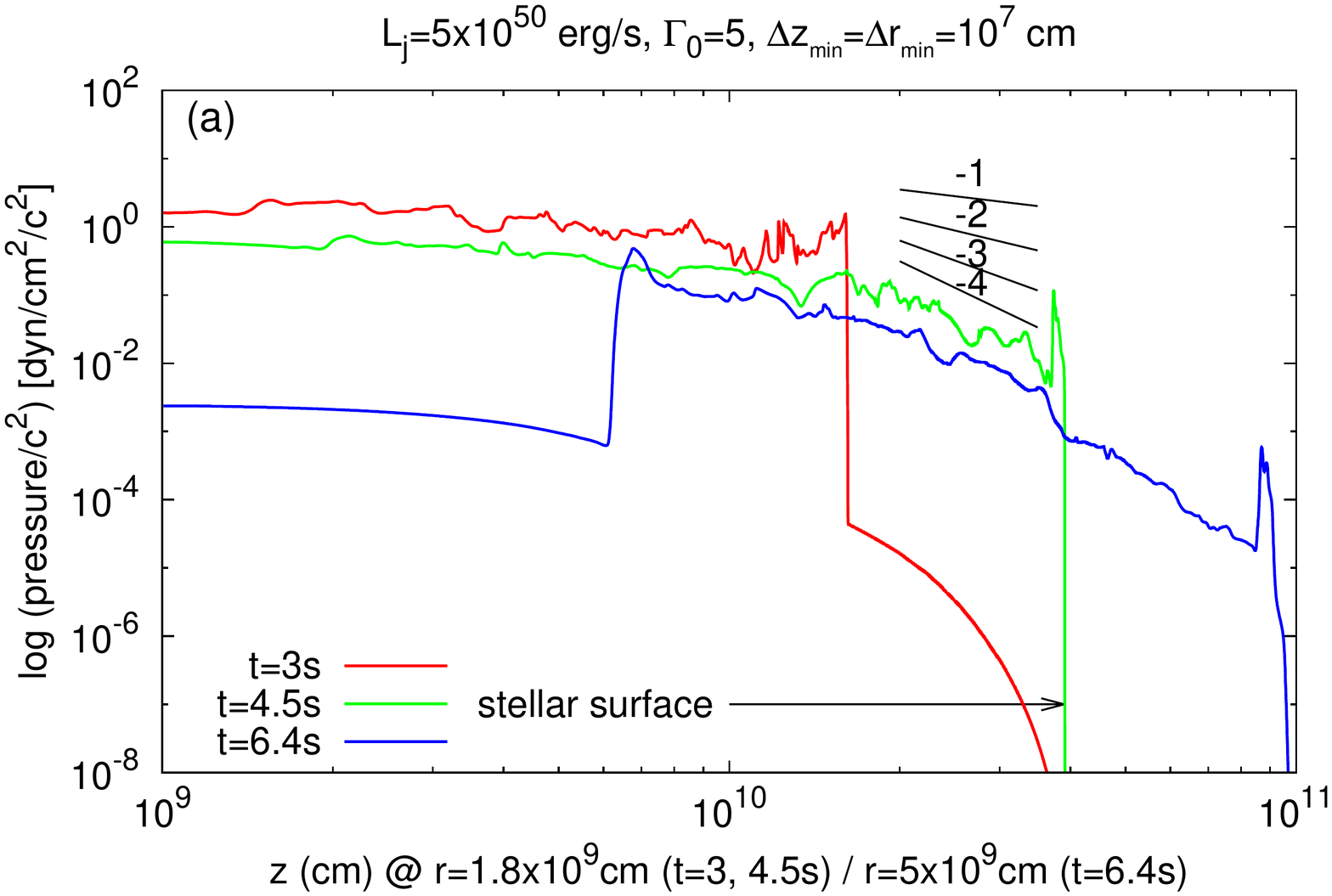}}\\
\rotatebox{0}{\plotone{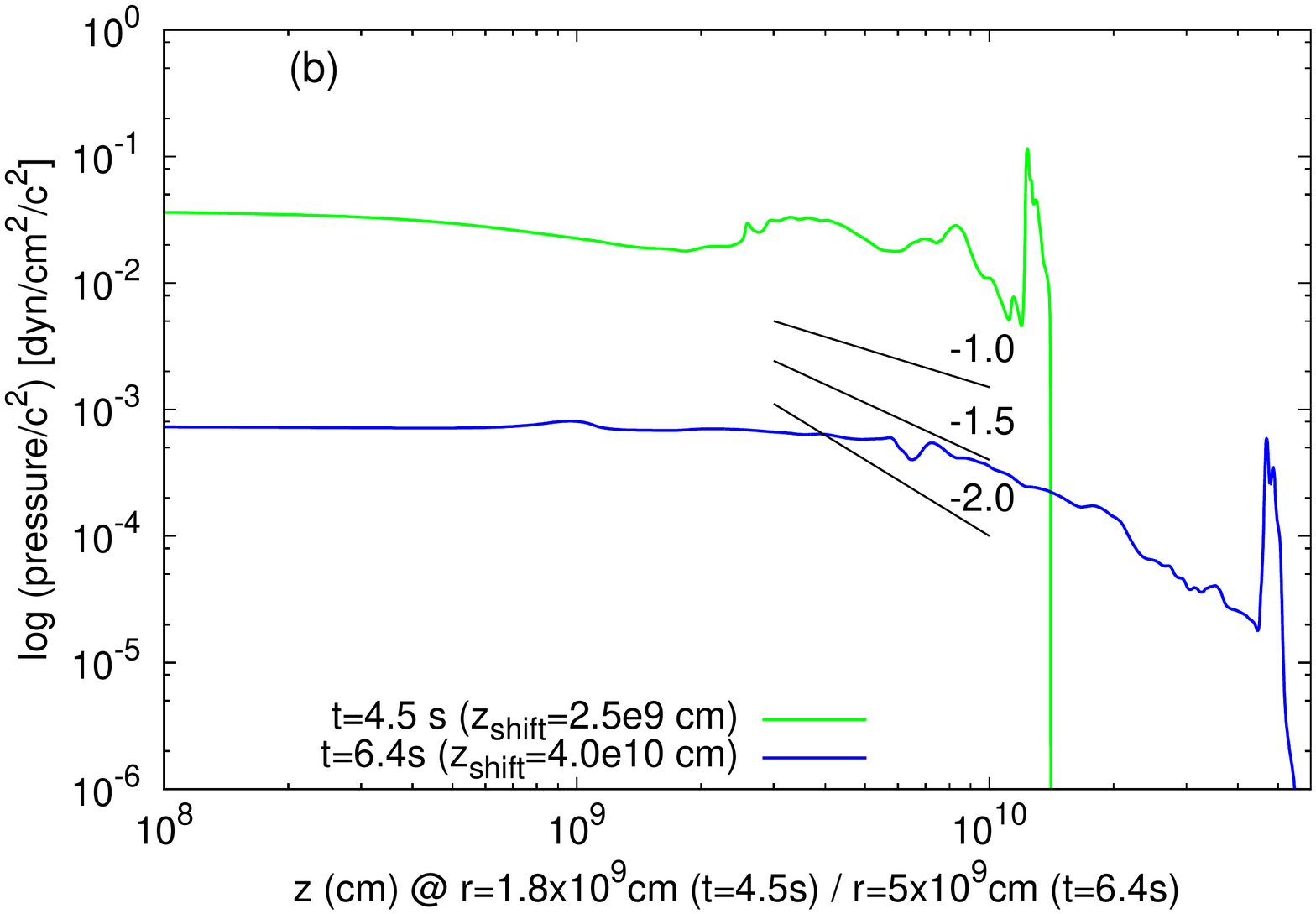}}\\
\rotatebox{0}{\plotone{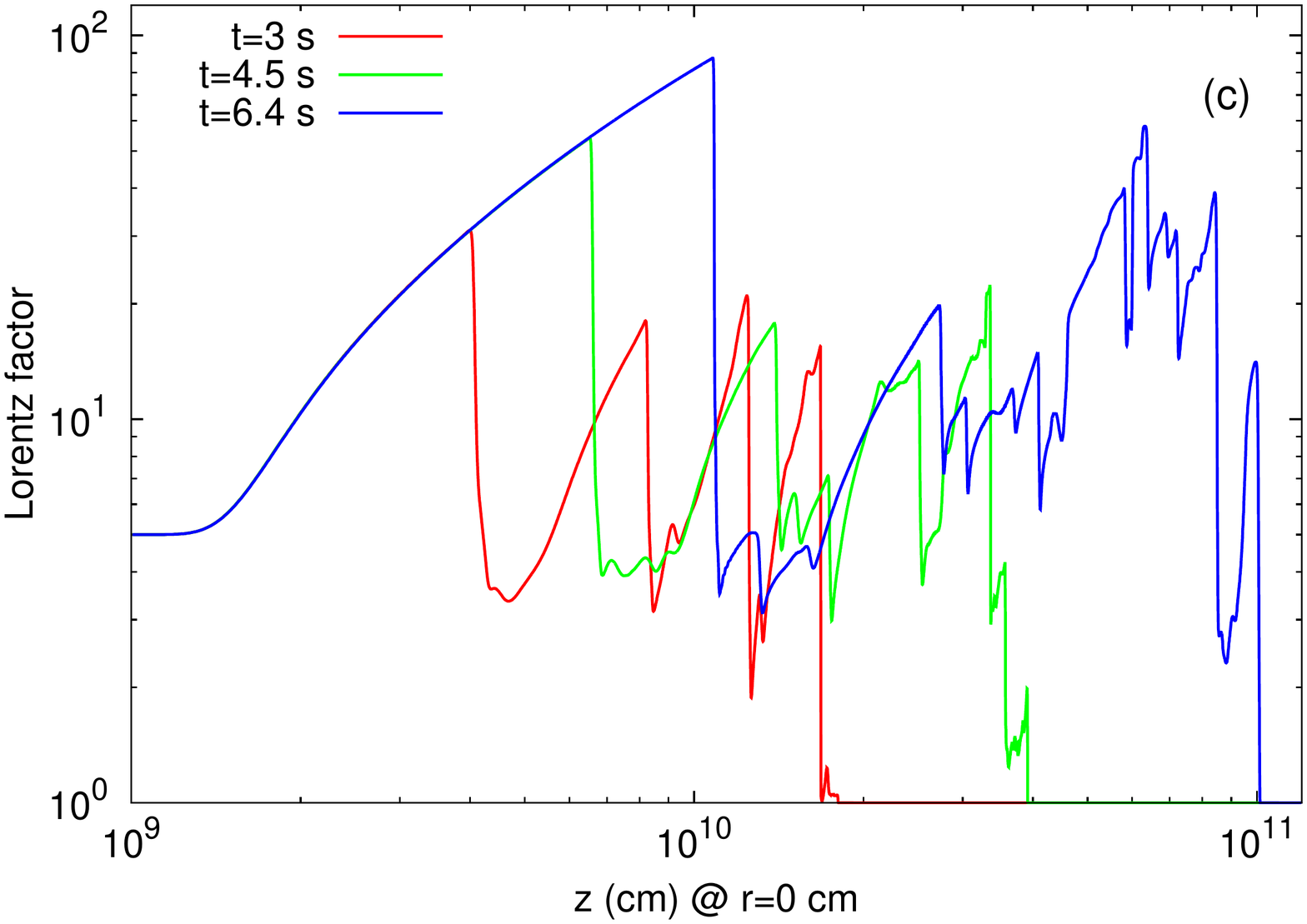}}
\caption{\label{fig:1Dpressure}
(a) One dimensional pressure profiles at 
$r=1.8\times 10^9$ cm and $t=3,~4.5$~s,
and at $r=5\times 10^9$ cm and $t=6.4$~s
for model G5.0.
The profile shows the cocoon region.
The shocked ambient gas appears as a thin shell at the jet head.
The pressure profile is almost homogeneous before the
jet breakout ($t=3$~s).
When the jet breakout occurs at $t=4.5$~s,
a pressure gradient can be seen around the jet head.
After the jet breakout at $t\ge 6.4$~s,
the pressure profile in the outer cocoon is about
$p\propto z^{-4}$.
(b) Same as (a) but only for $t=4.5~,6$~s
and the horizontal axis is $z-z_{\rm shift}$ to show
the pressure profiles measured from off-center origin
at $z_{\rm shift}$, which determines the evolution
of the jet expanding from the off-center around $\sim z_{\rm shift}$
for each time
($z_{\rm shift}=2.5\times 10^{9}\rm {cm}$ for
$t=4.5$~s and
$z_{\rm shift}=4\times 10^{10}\rm {cm}$ for $t=6.4$~s).
A pressure gradient can be seen around the jet head
at the time of the jet breakout ($t=4.5$~s)
and after the jet breakout ($t= 6.4$~s).
(c) One dimensional Lorentz factor profile along the $z$-axis
at $t=3,~4.5,$ and $6.4$~s for model G5.0.
}
\end{figure}

As shown in Figure~\ref{fig:Gam0_5_0045},
after the first collimation shock, 
the jet maintains an overall
cylindrical shape \citep{Bromberg11}.
The Lorentz factor increases to a few tens
and then drops to $\sim \Gamma_{0}$, i.e., the initial Lorentz factor,
after the first collimation shock, see Equation~(\ref{eq:gamma1}).
The Lorentz factor after the first collimation shock
remains constant with time on average,
although the pressure in the cocoon gradually decreases with time,
since the timescale for
the cocoon to change is much longer than that for the jet to cross the
cylindrical region.
Figure~\ref{fig:1Dpressure}(c)
shows the one-dimensional Lorentz factor profile along the $z$ axis.

We can understand the cylindrical evolution as follows.
The propagation velocity of the jet head
is comparable to the sound velocity of the shocked envelope (obviously)
and is much smaller than the sound velocity 
of the shocked jet (with high entropy) at the jet head.
The shocked jet goes backward around the jet 
and provides a constant cocoon pressure over the jet.
Namely, the flow in the cocoon is subsonic.
Although the mean sound velocity of the cocoon,
which is comparable with the transverse velocity of the bow shock,
is lower than the jet head velocity,
the matter is only partially mixed in the cocoon.
Thus, the gas in the cocoon can communicate with each other,
resulting in a homogeneous pressure profile in the cocoon.
Figure~\ref{fig:1Dpressure}(a) shows the pressure profile
at $r=1.8\times 10^9$~cm as a function of $z$.
At $t=3$~s (before the jet breakout),
we can see a relatively constant pressure profile
over the cocoon.
Since the jet is confined by the homogeneous pressure in the cocoon,
the jet maintains a cylindrical structure after passing the first
collimation shock.
As shown in Figure~\ref{fig:1Dpressure}(a) ($t=4.5~$s),
the homogeneous pressure decreases with time,
and the cylindrical radius of the jet increases with time.
The constant pressure profile in the cocoon
breaks around the head of the jet.

The Lorentz factor should be constant 
over the cylindrical jet because of flux conservation.
The Lorentz factor remains $\sim \Gamma_{0}$,
although some fluctuations appear in the Lorentz factor due to
the internal oblique shocks;
see Figure~\ref{fig:1Dpressure}(c).
The internal oblique shocks
occur in the cylindrical jet for several reasons.
The first reason is that the jet converges to the axis
after the first collimation shock.
The jet shrinks the cylindrical radius,
resulting in an oblique shock.
The second reason is similar to that for the first collimation shock.
The shrinking jet becomes over-pressured and bounces.
This is like the initial expansion at the injection and thereby 
leads to the second collimation shock.
Such a cycle of expansion and collimation is repeated several
times.
Note that the jet has a larger initial cylindrical radius
in the second and later collimations than in the first one.
The third reason is the fluctuation of the cocoon pressure \citep{Mizuta04}.
The cocoon is produced by the shocked jet and the shocked envelope.
The shocked jet at the jet head goes sideways
and also backward under the pressure of the shocked envelope \citep{Mizuta10}.
Together with the shear motion between the shocked jet and the shocked envelope,
a large vortex and a turbulent structure is formed in the cocoon.
The turbulence in the cocoon makes a perturbation
at the contact discontinuity between the cocoon and the jet,
which causes the oblique shocks in the jet.
Because shocks are tilted to the jet axis,
we call these shocks oblique shocks in the jet.

At the last stage of the propagation in the progenitor,
we identify a jet-breakout acceleration.
Since the density distribution of the envelope drops
exponentially at the stellar surface $R_*$~($z=4\times 10^{10}$
cm; see Figure~\ref{fig:radial_mass}),
the jet head advances rapidly ($\sim c$, the speed of light).
As a result, the pressure profile in the cocoon 
cannot remain constant around the jet head.
The pressure profile in the cocoon drops near the jet head;
see the one-dimensional pressure profile in the cocoon
in Figure~\ref{fig:1Dpressure}(a) at  $t=4.5$~s
when the jet breakout just occurs.
Figure~\ref{fig:1Dpressure}(b) shows the same as
Figure~\ref{fig:1Dpressure}(a) for but when the $z$-axis is shifted by
$z_{\rm  shift}$, where $z_{\rm  shift}$ is 
approximately the position at which the jet starts to expand.
The jet cannot maintain its cylindrical structure near the jet head
even after passing the oblique shock,
since the cocoon pressure drops before the jet breakout.
The jet expands in the decreasing cocoon pressure
and accelerates by converting thermal energy 
into kinetic energy.
The Lorentz factor of the jet increases to a few tens
even after the oblique shock.
See the Lorentz factor contour near 
the jet head in Figures~\ref{fig:Gam0_5_0045} and \ref{fig:1Dpressure}(c)
and the discussion in Section~\ref{sec:model.postbreak}.

\subsection{After the Jet Breakout}

Soon after the forward shock reaches the stellar surface,
jet breakout occurs.
The jet and the cocoon 
(a mix of the shocked jet and the shocked stellar envelope) start to expand
into the circumstellar matter which is assumed to be very dilute.
The expansion velocity of the cocoon is
comparable to the sound velocity 
before the jet breakout ($\sim$ a few tens of
percent of the speed of light).
So, the cocoon stays near the stellar surface for 10 s
(see Section~\ref{sec:duration}),
providing pressure for the jet confinement.
As shown in Figure~\ref{fig:1Dpressure}(b), for $t=6.4$~s,
the cocoon pressure is decreasing outward,
ranging from $P \sim {\rm const}$ to $P\propto (z-z_{\rm shift})^{-2}$
(the off-center case).

On the other hand,
the supersonic jet does not notice the cocoon profile
until a collimation shock is formed.
In the star, the jet is repeating a cycle of
the expansion and the (over-)collimation 
that maintains the cylindrical structure;
see Sections~\ref{sec:before} and \ref{sec:opening_angle}.
Near the
stellar surface, the jet also expands without noticing
the outwardly decreasing pressure before the shock.
Note that the expansion is off-center near the stellar surface,
not from the stellar center.
The off-center origin makes the pressure profile
shallower (Figure~\ref{fig:1Dpressure}(b))
than that shown in Figure~\ref{fig:1Dpressure}(a)
with a stellar center origin.
Since the pressure profile is still shallower than
$P \propto (z-z_{\rm shift})^{-2}$,
the jet is collimated
but the collimation is not enough to keep the cylindrical radius fixed
(see Section~\ref{sec:opening_angle}).
This expansion leads to the jet-breakout acceleration, 
as shown in Section~\ref{sec:model.postbreak}.
Even after the last collimating oblique shock,
a certain level of confinement continues
without forming a shock (or with a weak shock),
but the jet expands laterally, leading to
an additional jet-breakout acceleration 
(see Section~\ref{sec:model.postbreak}).
Finally the jet starts to expand freely
after the pressure profile gets steeper than
$P \propto (z-z_{\rm shift})^{-4}$.
As time goes on,
the steep pressure profile moves inward,
while the first collimation shock becomes large.
Then the jet starts a free expansion 
after crossing the first collimation shock.

\begin{figure}
\epsscale{1.7}
\plotone{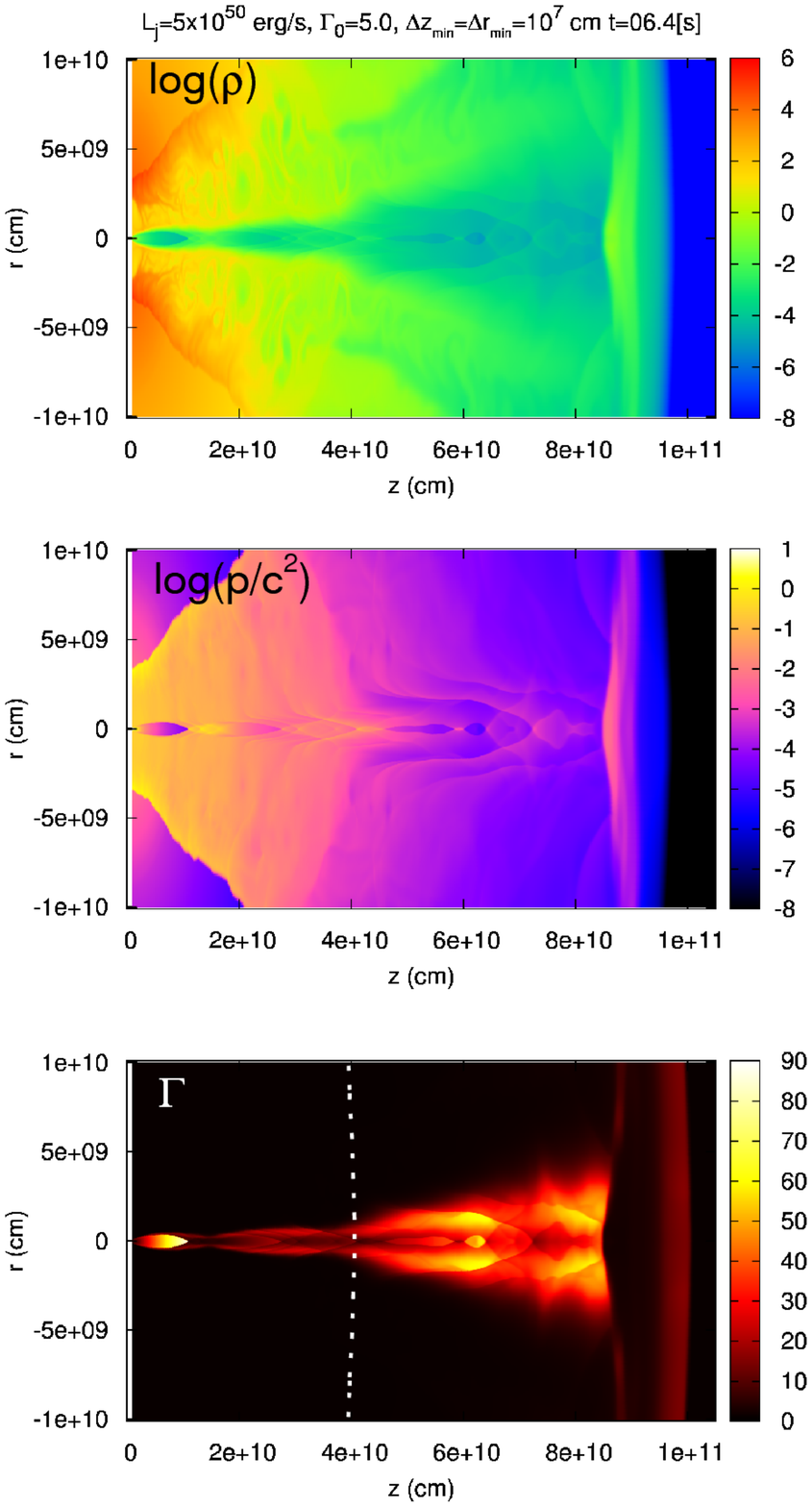}
\caption{\label{fig:Gam0_5_0139}
Same as Figure~\ref{fig:Gam0_5_0045} but 
after the jet breakout (at $t=6.4$~s,  model G5.0).
}
\end{figure}

Figure \ref{fig:Gam0_5_0139} shows
the mass density, the pressure, and the Lorentz factor contours
of the model G5.0 ($\Gamma_{0}=5$) at $t=6.4$ s.
There are many oblique shocks in the jet,
not only inside the progenitor but also outside the progenitor.
The oblique shocks outside the star are 
imprinted before the free expansion
and expand in a self-similar way,
because the internal shocks were not developed
in a freely accelerating flow without confinement \citep{KI+11}.
The jet advances with a velocity close to the speed of the light,
whereas the cocoon expands with a sub-relativistic speed.
As the cocoon pressure decreases,
the first collimation shock expands and
the converging point of the shock moves outside the progenitor star.

\subsection{Opening Angle After the Jet Breakout}
\label{sec:opening_angle}
\begin{figure*}
\epsscale{1.3}
\rotatebox{0}{\plotone{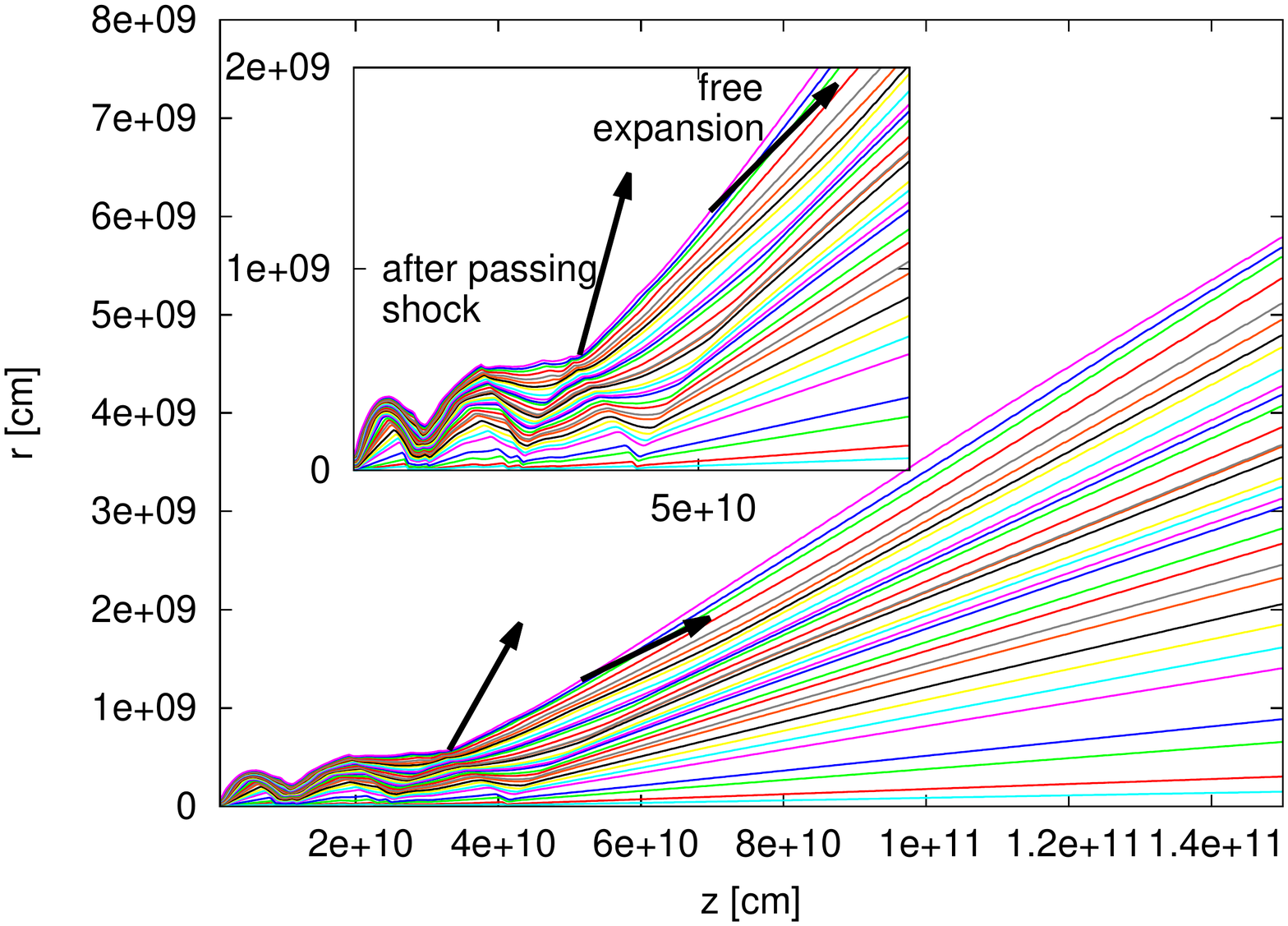}}
\caption{\label{fig:trajectory}
Particle trajectories injected with the jet
at the same time ($t=5$~s, model G5.0).
The arrows indicate the slope of the inverse of the
local Lorentz factor ($\Gamma^{-1}$)
at the time when the particles have passed the 
final collimating oblique shock before
 the free expansion (left arrow)
and at the time when the free expansion starts (right arrow)
for the laterally outermost particle.
The free expansion direction almost coincides with 
the slope of the inverse of the
local Lorentz factor ($\Gamma^{-1}$) when
the free expansion starts.
The inset displays a zoom in of  the range $1\times 10^9~{\rm cm}\le z \le
 8\times 10^{10}~{\rm cm}$.
Note that the aspect ratio of $z$ and $r$ is not unity.}
\end{figure*}

Since probe particles are introduced in the jet (Section \ref{sec:probe}),
we can trace the particle path and measure the opening angle
of the jet.
Figure \ref{fig:trajectory} shows
the traces of particles that are injected at the same time ($t=5$~s).
The path of each particle repeats the cycle
of expansion and collimation inside the star
and the last collimation near the stellar surface
is not so strong that the cylindrical radius
of the jet becomes large.
Just outside the progenitor ($R \simg R_*=4\times 10^{10}~{\rm cm}$),
the particles do not yet start a free expansion.
At some distance which depends on the particles,
the particle path finally becomes straight,
i.e., the jet freely expands as discussed in the previous section.
The position where the particle path becomes straight
depends on when and where the particle is injected.
The location of free expansion
starts at $\sim 5\times 10^{10}$~cm for early particles
which were near the head of the jet at the jet breakout.
Then, the locations to start the free expansion
move inward for subsequent particles.

The arrows in Figure~\ref{fig:trajectory} indicate
the slopes of the inverse of the local Lorentz factor
($\Gamma^{-1}$) for the laterally outermost particle.
The left arrow is $\Gamma^{-1}$ just after
the last collimating oblique shock ($z \sim ~3.3\times 10^{10}$~cm).
Although the jet behind the shock has a higher Lorentz factor 
than that in the star,
the arrow is pointing outside the jet opening angle.
Thus, the jet is still confined by the cocoon and can not expand freely,
drawing a concave particle path in Fig.~\ref{fig:trajectory}.
Finally, the path becomes straight from the base of the right arrow
($z\sim 5\times 10^{10}~{\rm cm}$).
The direction of the arrow almost coincides with
the freely expanding direction.
This means that the opening angle of the jet
is determined by the Lorentz factor of the flow when
the free expansion starts as in Equation~(\ref{eq:thjgab}).

The extrapolations of the free expansion lines
do not cross the center of the progenitor,
i.e., the explosion is off-center.
The explosion center moves gradually inward as time passes.
The off-center position is different 
even for particles injected at the same time. 

We note that the acceleration of
the Lorentz factor by a factor of $\sim 5$
does not take $\sim 5$ stellar radii.
This is because the expansion is
off-center and the initial size of the fireball is much smaller than
the stellar radius under the cocoon pressure.
The fireball expands by a factor of $\gtrsim 5$
before the free expansion, but the size of the fireball
is still comparable with the radius of the star $R_*$.
See also the discussion in Section~\ref{sec:model.postbreak}.

Since the particles with $h\Gamma < 100$
lose their potential to reach $\Gamma \ge 100$
by an adiabatic expansion,
only particles with $h\Gamma \ge 100$
contribute to the GRB prompt emission.
Baryon loading occurs for the particles
that are involved in the cocoon component before the free expansion,
because the turbulence in the cocoon mixes
the shocked jet and the shocked stellar envelope.
Baryon loading also occurs at the contact discontinuity
between the jet and the cocoon even after jet breakout
(mostly via numerical diffusion).
A certain level of baryon loading is unavoidable
through numerical diffusion
and we should be careful about it,
as discussed in Section~\ref{sec:resolution}.
Most particles that are injected at early times
exhibit $h\Gamma<100$ at large $z$.

\begin{figure}
\epsscale{1.3}
\rotatebox{0}{\plotone{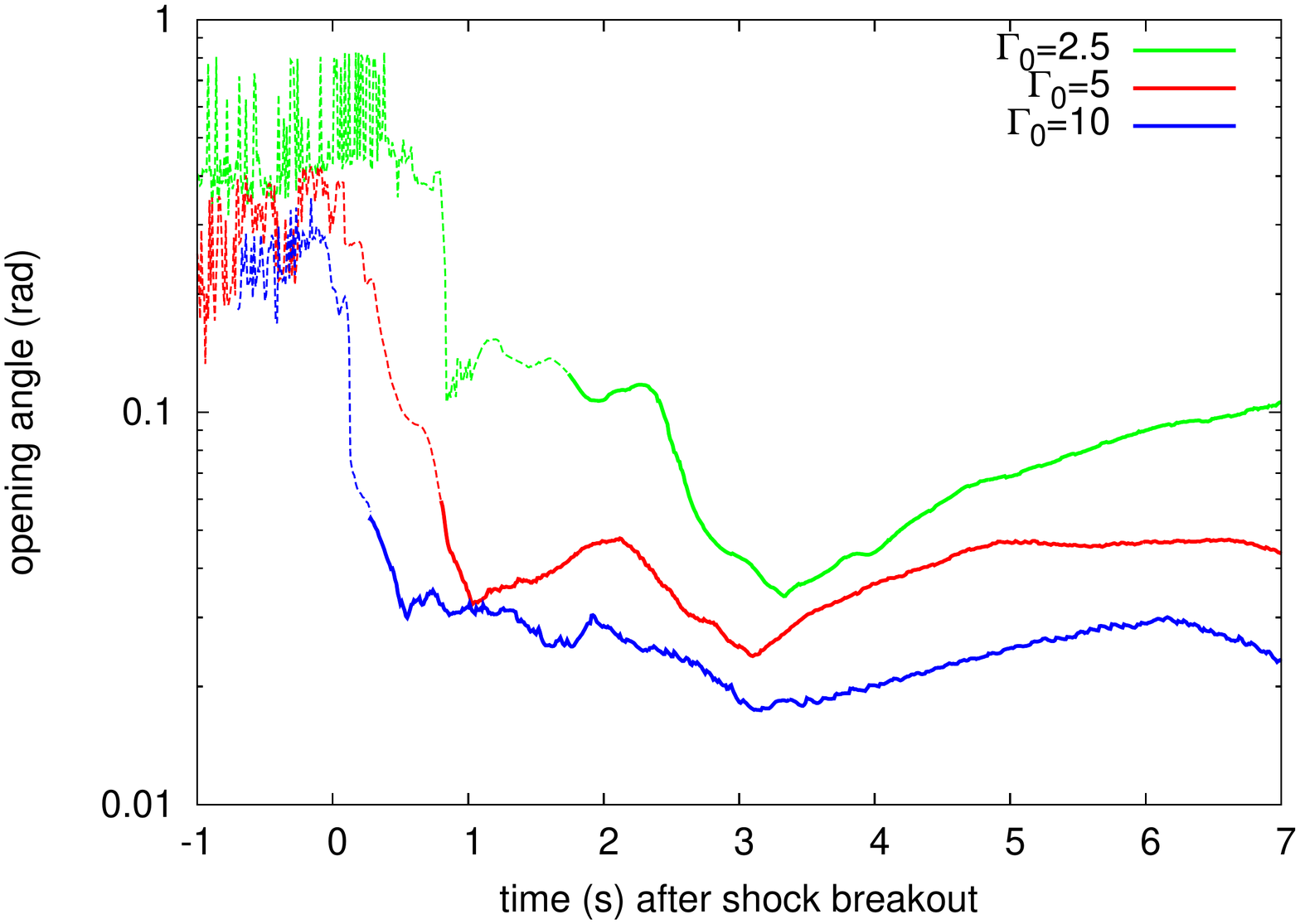}}
\caption{\label{fig:angle.time}
Opening angles as a function of time before and
after the shock breakout for the resolution
${\Delta z}_{\rm min}={\Delta r}_{\rm min}=10^7{\rm cm}$ cases
(models G2.5, G5.0, and G10).
The opening angles are measured by using probe particles
(see Section~\ref{sec:opening_angle}).
The components that satisfy $h\Gamma\ge 100$ are indicated
by the solid lines.
Those are components to be accelerated to $\Gamma \ge 100$
by the adiabatic expansion
and can contribute to the prompt emission.
The components with $h\Gamma < 100$ 
via the baryon loading from the shocked stellar envelopes
are shown by the dashed lines.
}
\end{figure}

We measure the jet opening angle $(\theta_j)$ as
the angle between the jet axis and 
the free expansion path for the laterally outermost particle.
Figure \ref{fig:angle.time} shows the time evolution of the jet opening
angle for each model.
The solid lines show the opening angles 
measured by the particles with $h\Gamma\ge 100$,
while the dashed lines correspond to $h\Gamma < 100$
that can not produce GRBs.
The time axis in Figure~\ref{fig:angle.time} 
corresponds to the time when the particles are injected.
Since the jet breakout time depends on the models,
we align the shock breakout time at $t=0$ in Figure~\ref{fig:angle.time}.
Thus the time $t<0$ indicates
that the particles move into the cocoon
before the shock breakout.

The opening angle of the jet is not constant at early times,
as shown in Figure~\ref{fig:angle.time}.
There are particles
with $h\Gamma < 100$ 
before and a few seconds after the shock breakout. 
Those particles are the components
that are engulfed into the cocoon before the jet breakout.
The cocoon is largely baryon-loaded
because the shocked envelope mixes with the shocked jet
by the shear interaction through the contact discontinuity.
Since the particles expand with the cocoon after the shock breakout,
those opening angles are relatively large.

The opening angle drops after the shock breakout in Figure~\ref{fig:angle.time}.
At the same time, the baryon-poor GRB component $h\Gamma\ge 100$ appears.
We note that our results are not sensitive to
the threshold value of $h\Gamma\ge 100$
because the baryon-loaded flow and the baryon-poor flow are well separated,
irrespective of the uncertainty in the numerical baryon loading
(see also Section~\ref{sec:angular}).
The opening angle settles down to a nearly constant value
a few seconds after the shock breakout.
Small time variations of the opening angle
are caused by the fluctuations of the local Lorentz factor
just before the free expansion, according to Equation~(\ref{eq:thjgab}).
The local Lorentz factor depends on 
where and how the particle crosses the collimation shock.
Since the size of the first collimation shock grows 
to a size comparable with the stellar radius,
the later particles only pass the first collimation shock 
before the free expansion.

\begin{figure}
\epsscale{1.3}
\rotatebox{0}{\plotone{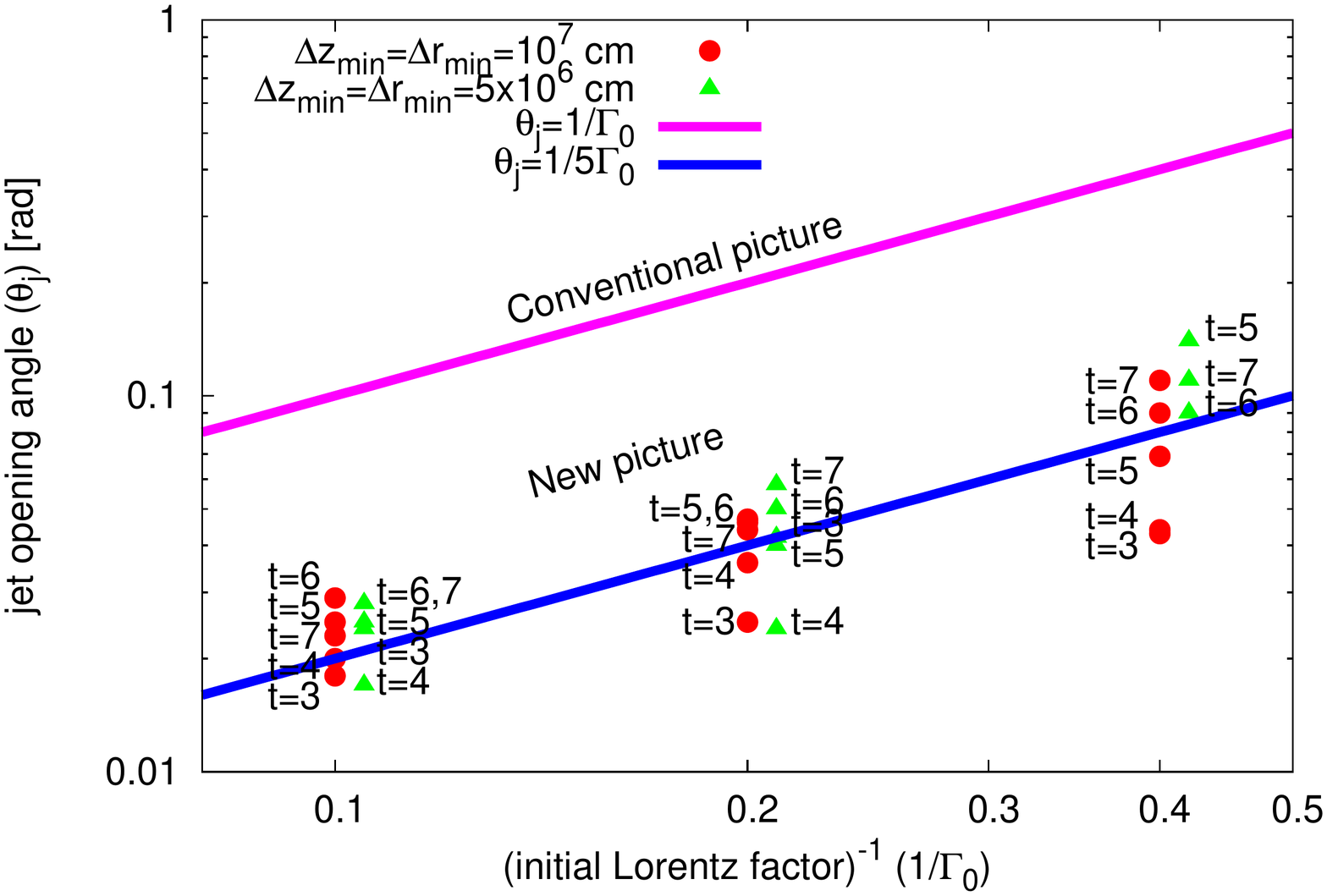}}
\caption{\label{fig:angle.gamma}
Opening angle ($\theta_j$) vs. the inverse of the initial Lorentz factor ($\Gamma_{0}^{-1}$).
The opening angles at $t=3,~4,~5,~6$, and 7~s are shown
for models G2.5, G5.0, and G10 with the resolution ${\Delta z}_{\rm
 min}={\Delta r}_{\rm min}=10^7$~cm (red circles).
The opening angles are also shown
at $t=3,~4,~5,~6$, and 7~s for models G5.0H and G10H
and at $t=5,~6,$ and 7~s for model G2.5H
with the resolution ${\Delta z}_{\rm min}={\Delta r}_{\rm min}=5\times
 10^6$~cm
(green triangles ; the green triangles are
slightly shifted to the right).
We also plot lines of $\theta_j=\Gamma_{0}^{-1}$ (the conventional picture
based on the naive expectation) and
$\theta_j=(5\Gamma_{0})^{-1}$ (our new picture obtained
by numerical simulations).
}
\end{figure}

Most importantly,
the opening angles 
in Figure~\ref{fig:angle.time}
are much smaller than $\Gamma_{0}^{-1}$,
contrary to our naive expectations
($1/\Gamma_{0} \sim 0.2$ rad for $\Gamma_{0}=5$ and
$1/\Gamma_{0} \sim 0.1$ rad for $\Gamma_{0}=10$).
Figure \ref{fig:angle.gamma} shows the $\theta_j-\Gamma_{0}^{-1}$ plot
for the models.
We plot the opening angles at different times (from $t=3~$s to 7~s)
since they fluctuate, as shown in Figure~\ref{fig:angle.time}.
The opening angles of the jet are not on the line 
$\theta_j \sim \Gamma_{0}^{-1}$
but on the line $\theta_j \sim (5\Gamma_{0})^{-1}$.
This is our main result;
the opening angle of the GRB jet from collapsars
for the first 10 seconds (see Section~\ref{sec:duration})
is roughly given by
\begin{eqnarray}
\theta_j \simeq {1\over 5\Gamma_{0}}.
\label{eq:thj}
\end{eqnarray}
If the activity of the engine continues,
the opening angle of the jet will increase
as shown by \citet{Morsony07}.

\subsection{Angular Profile of the Jet After the Jet Breakout}
\label{sec:angular}
\begin{figure}
\epsscale{1.3}
\rotatebox{0}{\plotone{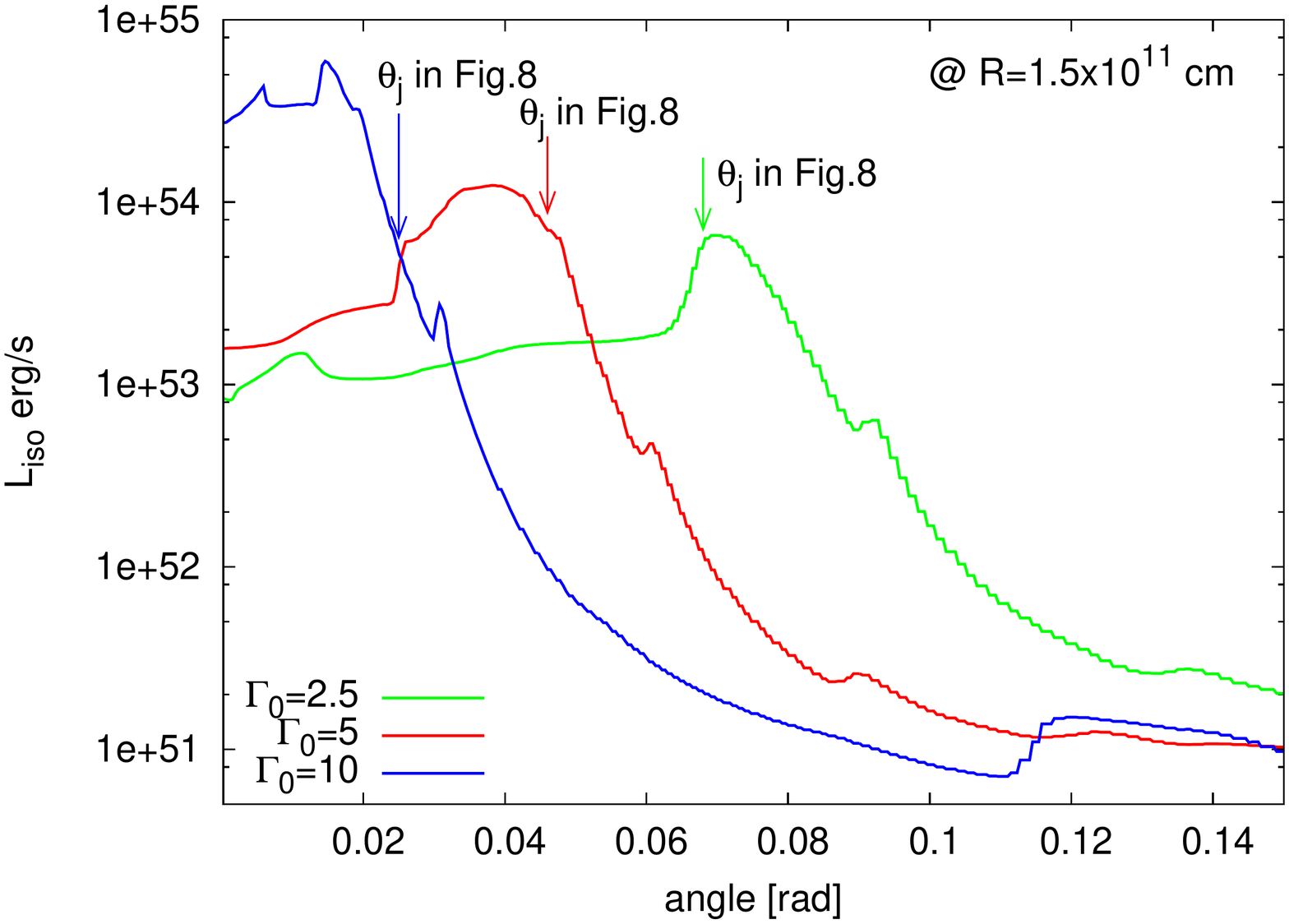}}
\caption{\label{fig:angular}
Angular energy density distribution after the jet breakout
at a spherical radius ($R=1.5\times 10^{11}~{\rm cm}$) for models
G2.5 at $t=14.8$~s (green), G5.0 at $t=12.5$ (red), and G10 at
 $t=11.9$~s (blue).
The time is when the forward
shock reaches $z=3 \times 10^{11}~{\rm cm}$.
The energy density distribution of the jet shows a hollow-cone
structure.
The jet opening angles at $t=5$~s
in Figure~\ref{fig:angle.time} are indicated by the arrows for each model,
and coincides with 
the position at which the energy density starts to drop exponentially.
}
\end{figure}

Figure \ref{fig:angular} shows the angular distribution of isotropic
luminosity
($L_{\rm iso}$) 
using the hydrodynamic quantities of the fluid calculations
without any constrain on $h\Gamma$,
at $R_{a}=1.5\times 10^{11}~{\rm cm}$
at the time
when the forward shock reaches $z_{\rm FS}=3 \times 10^{11}~{\rm cm}$,
for models G2.5 ($t=14.8$~s), G5.0 ($t=12.5$), and G10 ($t=11.9$~s).
The angle is in spherical coordinates with the origin at the center
of the progenitor.
The angular distribution of the isotropic luminosity
profile shown in Figure~\ref{fig:angular}
roughly corresponds to the particles
at $t=(z_{\rm FS}-R_{a})/c=5~{\rm s}$ after the shock breakout
in Figure~\ref{fig:angle.time}.
The arrows indicate the opening angle of the jet
measured in Figure~\ref{fig:angle.time}
for each model.
Each arrow roughly coincides with the rim
at which the isotropic luminosity starts to drop exponentially.
Our results are not sensitive to
the threshold value of $h\Gamma\ge 100$,
irrespective of the uncertainty in the numerical baryon loading.

Figure~\ref{fig:angular} shows a hollow-cone jet structure.
The angular distribution of the isotropic luminosity is high at the rim
and drops exponentially at the edge.
Then it gradually decreases at large angles.
The high isotropic luminosity rim part
is produced by the shock between the expanding jet and
the high density cocoon before the free expansion.
The other oblique shocks produced in the star
are also imprinted on the jet structure even after the jet breakout;
see Figure~\ref{fig:Gam0_5_0139} and \ref{fig:angle.time}.

Angular distributions of the jet are also shown by 
\citet{Zhang04} \citet{Morsony07}, and \citet{Mizuta09}.
Since these authors take the radial integration of the energy density
or the time integration of the energy flux at a certain radius,
as opposed to a snapshot of the isotropic luminosity,
we can not simply compare their results with ours.
In fact, there are some differences,
but it is difficult to identify the reasons
for these diffusions (numerical diffusion, the initial jet size etc.).

\subsection{The $\Gamma_{0}=2.5$ Case}
\begin{figure}
\epsscale{1.7}
\plotone{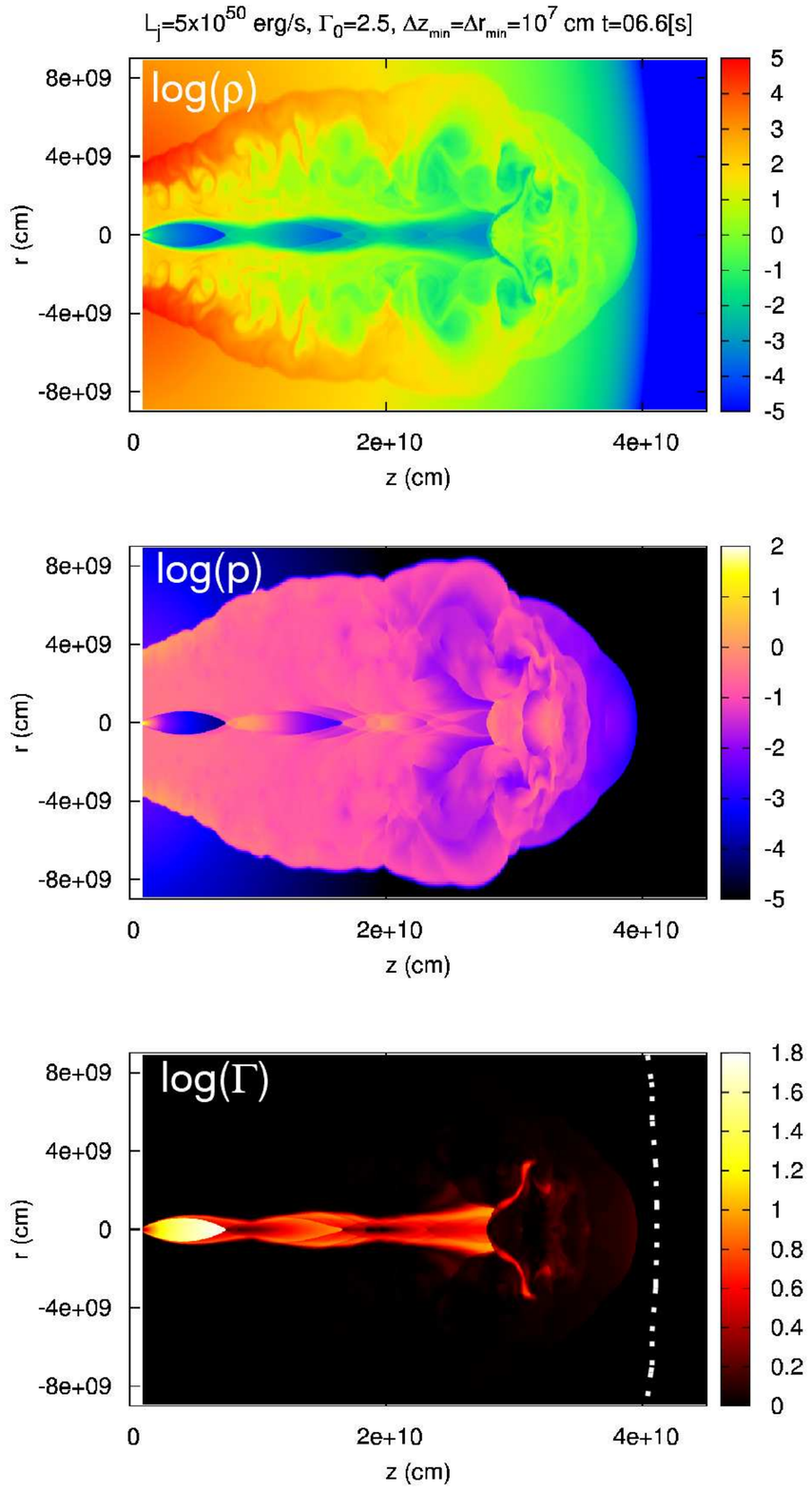}
\caption{\label{fig:Gam0_25_0066}
Same as Figure~\ref{fig:Gam0_5_0045} but $\Gamma_0=2.5$ (model G2.5), 
just before the shock breakout ($t=6.6$~s).
}
\end{figure}

The behavior of the jet for the case $\Gamma_{0}=2.5$
is somewhat different from other cases
($\Gamma_{0}=5$ and 10).
The jet opening angle 
at the injection point for the model $\Gamma_{0}=2.5$
is larger than those for other models,
since the initial opening angle is $\sim \Gamma_{0}^{-1}$.
The cylindrical radius of the jet
becomes larger than those of other models
(see the analytic study of the
jet dynamics in the progenitor in Section~\ref{sec:model.prebreak}).
As the cylindrical radius of the jet
increases before the jet breakout,
the momentum flux per area pushing the stellar envelopes decreases,
resulting in a strong reverse shock.
As a result, the forward shock and
the reverse shock go away from each other, as shown 
in Figure~\ref{fig:Gam0_25_0066}.
The reverse shock is far from
the progenitor surface at $z\sim 2.4\times 10^{10}$~cm
at the time when the forward shock reaches the stellar surface,
$z\sim 4\times 10^{10}$~cm.
Since the cylindrical radius of the jet is very large,
a large fraction of the shocked jet and the shocked envelope
can not go sideways and remains in the jet head.
The mass is collected at the jet head like a snowplow or a plug
\citep{Zhang04,Mizuta06}.
Even after the shock breakout,
the plug remains on the axis
and affects the jet advance for a while.
As the plug moves away from the star,
the jet can go around the plug more easily
and the effect of the plug decreases.
A similar structure would also appear for low-luminosity jets 
with $L_j < 5\times 10^{50}~{\rm erg ~s}^{-1}$
\cite[e.g.,][]{Toma07},
while the plug would be reduced for a non-axisymmetric case
\citep{Zhang04}.
We will study the luminosity dependence in the near future.

The plug affects
the time evolution 
of the opening angle after the jet breakout.
The opening angle in the case of $\Gamma_{0}=2.5$
does not drop quickly, compared with the other models.
It takes a few seconds to start to drop
and produces a much smaller angle than $\Gamma_{0}^{-1}$.
As shown in Figure~\ref{fig:Gam0_25_0066},
there is a baryon-rich plug ahead of the jet
when the jet breakout occurs.
The jet is scattered by the plug for a few seconds
after the jet breakout.
As a result the opening angle of the jet is large for a while.
The following jet,
which moves far away from the progenitor and hence does not interact
with the plug, has a small opening angle, as seen in our other models.

\subsection{Resolution Study}
\label{sec:resolution}
\begin{figure}
\epsscale{1.1}
\plotone{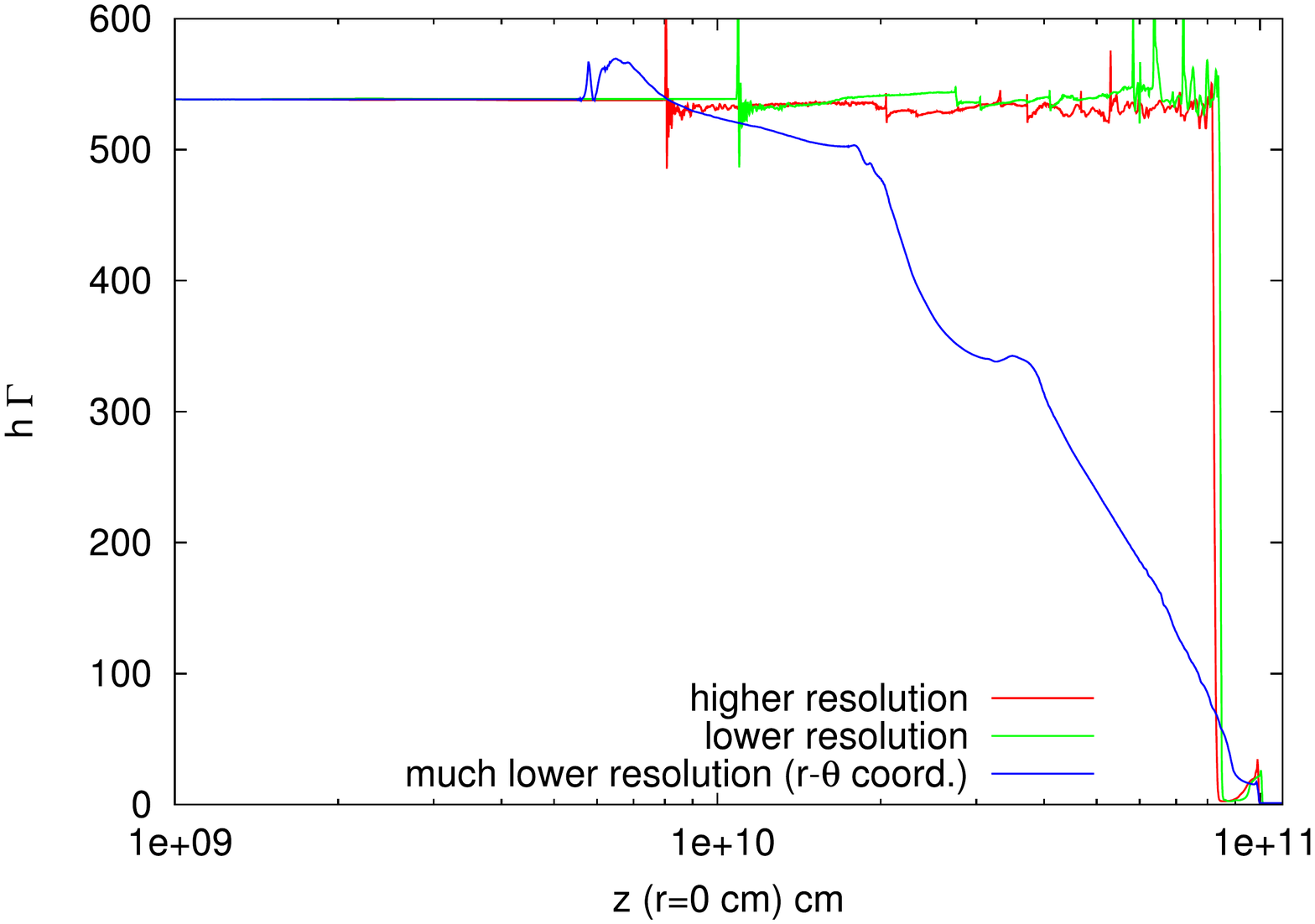}
\caption{\label{fig:1Dhgamma}
One dimensional profile of 
Bernoulli's constant
$h\Gamma$ along the jet axis
for $\Gamma_{0}=5$ at $t=6.4$~s
when the forward shock reaches $z \sim 10^{11}~{\rm cm}$.
Both resolution cases, i.e.,
${\Delta z}_{\rm min}={\Delta r}_{\rm min}=10^7{\rm cm}$
(model G5.0)
and
${\Delta z}_{\rm min}={\Delta r}_{\rm min}=5\times 10^6{\rm cm}$
(model G5.0H)
are shown.
$h\Gamma$ is conserved up to the jet head
except for some fluctuations in the internal shocks.
On the other hand, 
$h\Gamma$ drops near the jet head
for a simulation with similar jet parameters
but low-resolution spherical coordinates,
$\Delta \theta=0.^{\circ}25$, \citep{Mizuta11}
due to baryon loading via the interaction with the stellar envelope.
}
\end{figure}

We completed hydrodynamic simulations with
higher resolution as discussed in Section \ref{subsec:grid}.
The initial jet contains a small amount of baryons,
while the stellar envelope contains abundant baryons,
i.e., $h\Gamma\sim 1$.
As the jet proceeds in the stellar envelope,
a contact discontinuity is formed
between the shocked jet and the shocked
stellar envelope.
Our numerical calculations use fixed grid points.
Artificial baryon loading could happen when the discontinuity
crosses the grids via numerical diffusion.
Once the gas is polluted by baryons ($h\Gamma \sim 1$),
it is not accelerated to a large Lorentz factor
by the adiabatic expansion. 
Figure \ref{fig:1Dhgamma} shows the
one-dimensional profile of $h\Gamma$ along the jet axis
for $\Gamma_{0}=5$ at $t=6.4$~s
when the forward shock reaches $z \sim 10^{11}~{\rm cm}$
with both resolution cases, i.e.,
${\Delta z}_{\rm min}={\Delta r}_{\rm min}=10^7{\rm cm}$
and
${\Delta z}_{\rm min}={\Delta r}_{\rm min}=5\times 10^6{\rm cm}$.
In both cases, Bernoulli's constant $h\Gamma$ is conserved 
all the way to the jet head
except for some fluctuations at the internal shocks.
If we use poor resolution, 
Bernoulli's constant $h\Gamma$ is not well conserved.
For example, the case of \citet{Mizuta11} is also shown.
$h\Gamma$ is conserved up to a half of the jet
but not near the jet head due to the numerical baryon loading.
\citet{Mizuta11} adopted similar jet parameters
but a spherical coordinate for hydrodynamic simulations
with $\Delta \theta=0.^{\circ}25$ around the jet axis.
This is one of the reasons for using much higher resolution grid points
for detailed quantitative discussions on the opening angle of the jet.

Figure~\ref{fig:angle.gamma} shows the opening angles
for different resolutions.
The results are similar between different resolutions.
Therefore the relation $\theta_j \sim (5\Gamma_{0})^{-1}$ 
in Equation~(\ref{eq:thj}) seems robust.

\begin{figure}
\epsscale{1.7}
\plotone{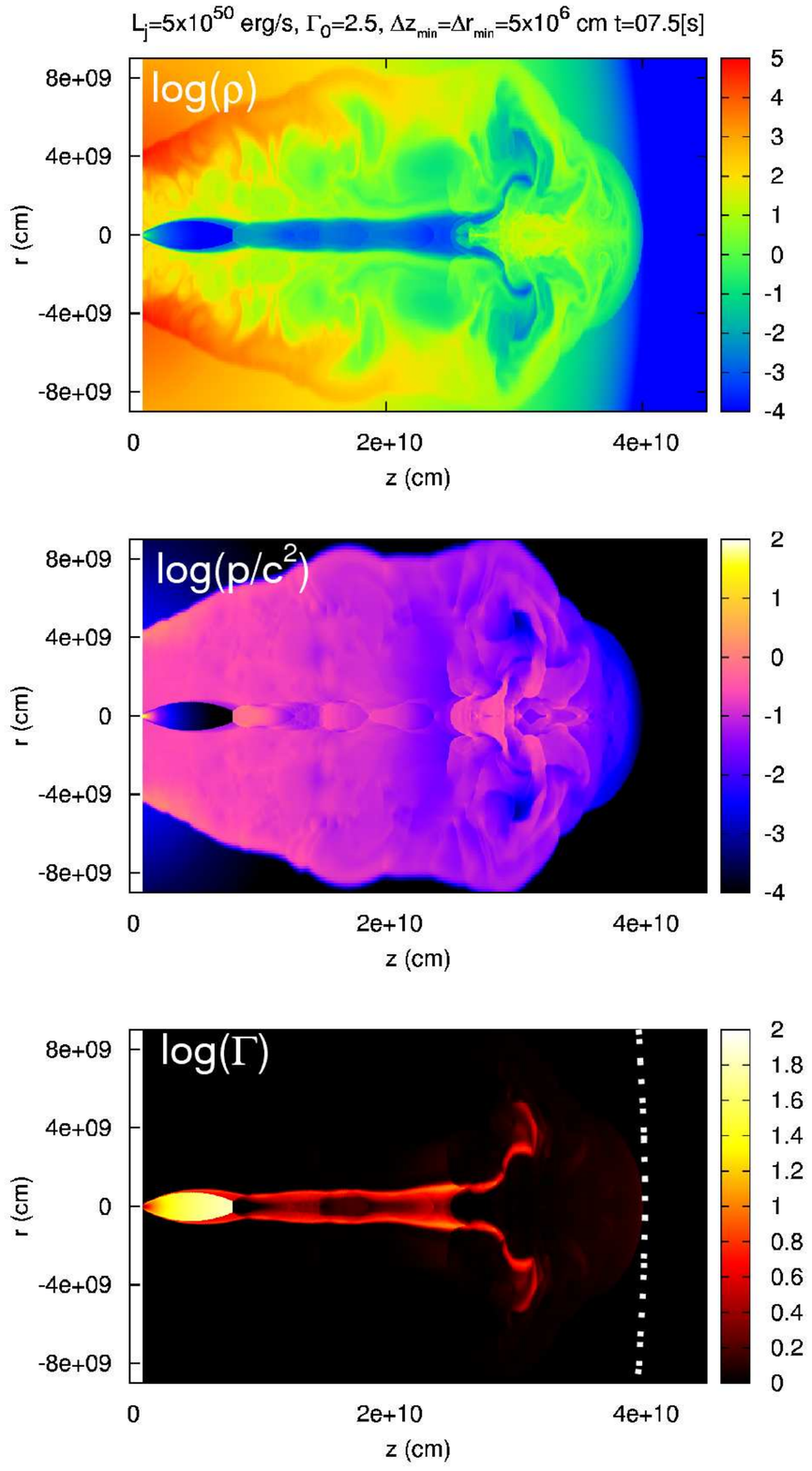}
\caption{\label{fig:resolution}
Same as Figure~\ref{fig:Gam0_5_0045} but $\Gamma_{0}=2.5$ 
with high resolution (model G2.5H), 
just before the shock breakout ($t=7.5$~s).
}
\end{figure}

Although the differences are small
between the models  G2.5 and G2.5H  ($\Gamma_{0}=2.5$), 
they depend on the resolution
more sensitively than the other models.
Figure \ref{fig:resolution} shows the mass density,
the pressure, and the Lorentz factor contours
of the model $\Gamma_{0}=2.5$ with the high resolution
at $t=7.5~$s when the shock breakout occurs.
The shock breakout times (6.6 s and 7.5 s)
are different by $\sim 1$ s
even if
the jet parameters are the same in Figure\ref{fig:Gam0_25_0066}
and \ref{fig:resolution}.
The high resolution case takes about 1.2 times longer than
the lower resolution case.
One should note that the recollimation shocks are different.
The low resolution one converges on the axis while
the high resolution one makes a Mach disk.
The difference is probably caused by the nonlinear evolution
of the jet and the cocoon dynamics.
A small difference in the oblique shocks near the head of the jet
changes the cross-sectional radius of the reverse shock
and hence the jet head speed.
This produces the difference
in the appearance of the vortices 
and the turbulence in the cocoon \citep{Mizuta10,Morsony10}.
The turbulence itself is also a non-linear process.
The numerical diffusion of the baryon loading would also alter the dynamics.
The size of the head plug is also larger for the high resolution case
than that for the low resolution case,
when the jet breakout occurs.

\subsection{Initial Jet Size}
\label{sec:initail_size}
At the injection point $z_{\min}=10^9$~cm,
we set the initial cylindrical radius of the jet
to be $r_0=8\times 10^7~{\rm cm}$.
The initial cylindrical radius should be smaller
than the value in Equation~(\ref{eq:rj}),
the cylindrical radius of the jet
after the collimation shock.
Otherwise, the jet evolution is different.
The jet does not show an initial expansion
but maintains its the incorrect cylindrical radius.
This gives wrong dynamics for
the breakout time, the cocoon energetics, and the opening angle.
That is why we start the numerical simulation
with so small an initial cylindrical radius.
At least several or 10 grid points are necessary
for covering the initial cylindrical radius
because we have to resolve the internal shocks.
This sets the scale for the fine grid points
as used in this study.
Our simulations use one of the finest resolutions so far.

\subsection{Comparison with Previous Studies}
We have run one of the highest resolution 
hydrodynamic simulations of jet propagation.
High resolution reduces the numerical baryon loading
which affects the jet dynamics.
Since the numerical baryon loading is the most dangerous
at the boundary with a high contrast of the enthalpy $h$,
the highest resolution grid points appear around the region
that covers the jet and some part of the cocoon.
The highest resolution grid size is comparable to or better
than that used in \citet{Morsony07}, and \citet{Lazzati13}.
The highest resolution region is much larger than
that used in \citet{Morsony07}.

Since we have run quite high resolution calculations
for hydrodynamic simulations of the jet dynamics,
the computational domain is restricted to about 10 times larger
than the progenitor size.
This is smaller than that
used in \citet{Mizuta11,Nagakura11}, and \citet{Suzuki13}.
Our discussion is restricted to only several seconds
after the jet breakout. 
The computational box size is comparable with
that used in \citet{Morsony07} who discussed
the time evolution of the opening angle of the jet.
The requirement that the initial cylindrical radius of the jet
should be sufficiently small is
also one of the reasons for using high-resolution grid points
(see Section~\ref{sec:initail_size}).
We have paid special care by applying
a sufficiently small cylindrical radius of the initial jet.
The dynamics of our results may be different from
those of \citet{Morsony07}, who took
the initial radius to $\sim 1.76\times 10^8$~cm
for the $\theta_0=10^{\circ}$ case.
This is about 2.2 times larger than our initial cylindrical radius of the jet
($r_0=8\times 10^{7}$~cm).

Our numerical results show that the opening angle of the jet
after jet breakout is narrower than 
what we expect ($\theta_j\sim 1/\Gamma_{0}$).
This trend is consistent with previous numerical simulations,
for example, \citet{Morsony07}, \citet{Lazzati09},
\citet{Mizuta11}, and \citet{Lazzati13}
who pointed out that the opening angle after the jet breakout
is smaller than the initial opening angle ($\theta_0$).
In this paper we introduce much more sophisticated
analysis with probe particles that allows
us to follow the Lagrangian motion of each element.
We determine the correlation between the initial Lorentz factor
and the opening angle of the jet, i.e., $\theta_j\sim 1/5\Gamma_0$.
We also find where and how the gas starts free expansion.
The position where expansion starts is 
off-center and also different for each particle
around $\sim 4 \times 10^{10}$ cm.
We identify that the jet breakout acceleration occurs and
that the local Lorentz factor of the flow just before the free expansion
determines the opening angle of the jet.
Since we measure the opening angle based on 
the free expansion direction with respect to the jet axis,
the angle is different from that measured by \citet{Morsony07},
who defined the jet opening angle by measuring how the
GRB jet component ($\Gamma_{\inf}\equiv h\Gamma \ge\Gamma_{cr} $)
spreads out at $R=1.2\times 10^{11}$~cm in spherical coordinates
(the progenitor center is the origin of the coordinate).
The different measurement methods result in about a 30 \%
difference in the opening angle of the jet.

\section{ANALYTIC MODELS}
\label{sec:analytic}
Let us consider analytical models of the jet propagation
and the dynamics of the opening angle.
The evolution is generally divided into two phases, i.e., 
before and after the jet breakout.

In Section~\ref{sec:model.prebreak}, we discuss the pre-breakout phase.
\citet{Bromberg11} investigated the pre-breakout evolution in detail.
Before the breakout, the jet is inside the progenitor star
and collimated by the cocoon pressure.
Since the cocoon pressure is constant (subsonic inside),
the jet becomes cylindrical through the first collimation shock.
The cylindrical (and stationary) flow
has a constant Lorentz factor because of flux conservation.
If the Lorentz factor were constant even at the jet breakout,
the opening angle of the jet would be the inverse of the initial
Lorentz factor (or the initial opening angle; see below),
as shown in Figure~\ref{fig:break},
but this is not the case as shown in Section~\ref{sec:model.postbreak}.

\begin{figure}
\epsscale{1.}
\plotone{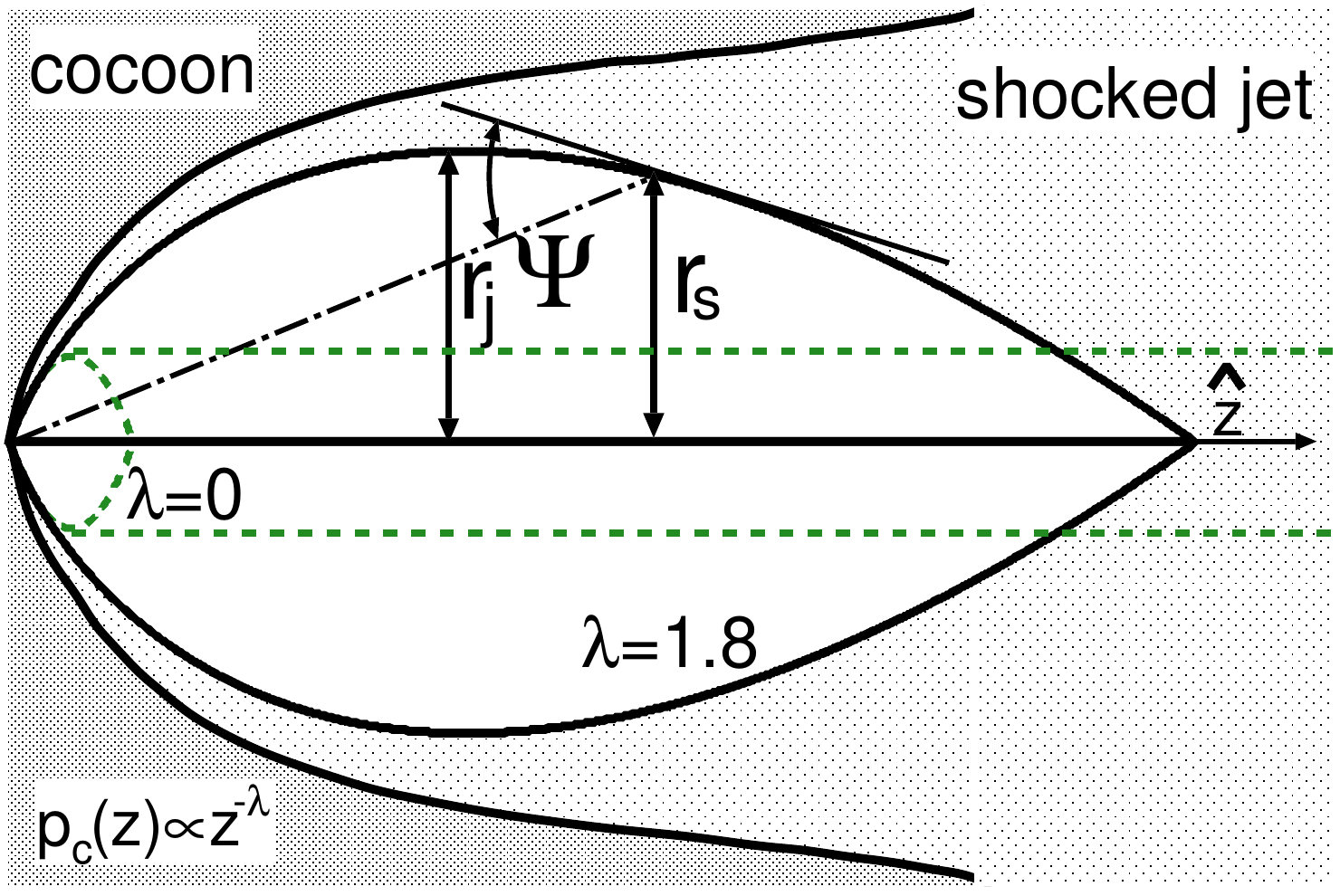}
\caption{\label{fig:jet_schematic}
Structure of the jet and the collimation shock. 
The collimation shock appears
by the interaction between the expanding jet
and the high pressure cocoon.
The collimation shock converges at $\hat z$.
Solid lines show
the collimation shock and the jet structure
after the collimation shock 
for a decreasing pressure case ($\lambda =1.8$),
whereas green the dashed lines show
the collimation shock and the cylindrical
 jet structure after the collimation shock 
for the constant pressure case ($\lambda =0$).
}
\end{figure}

We present analytical formulae of the jet propagation
for easy comparison with our numerical simulations.
We also calibrate  the model parameters,
taking great care with the baryon contamination and the initial jet size.
We compare the evolution of three physical quantities,
the jet head position, the jet cylindrical radius, 
and the collimation position, with numerical calculations.
The model has one free parameter $\eta$ in Equation~(\ref{eq:pc}), 
which is obtained by fitting the numerical results.

In Section~\ref{sec:model.postbreak},
we model the post-breakout phase.
After the breakout, the cocoon pressure is not constant
but decreases outward (see Figure~\ref{fig:1Dpressure}).
This leads to less collimation, a wider jet, and hence a larger Lorentz factor.
Therefore, the jet-breakout acceleration occurs almost inevitably
if the external medium has a finite size.
We estimate that the jet-breakout acceleration boosts
the Lorentz factor of the jet by a factor of several ($\sim 5$).
This is the reason why the naive picture in
Figure~\ref{fig:break} is not correct
but the jet opening angle becomes $\sim 1/5\Gamma$,
as shown numerically
in Section~\ref{sec:opening_angle} and Figure~\ref{fig:angle.time}.

\subsection{Jet Evolution in Constant External Pressure}
\label{sec:model.prebreak}

First we consider the jet evolution inside the progenitor star.
\citet{Bromberg11} provided a detail analysis.
Here we make the analytical formulae easier to use
in the calibration of the model parameters.
The jet dynamics is controlled by three processes:
(1) the jet head, (2) the cocoon, and (3) the collimation.
After combining these dynamics, 
we can describe the evolution of 
the jet head position by Equation~(\ref{eq:zh2}), 
the jet cylindrical radius by Equation~(\ref{eq:rj}), 
and the collimation position by Equation~(\ref{eq:hatz}).
There is one free parameter $\eta$ by Equation~(\ref{eq:pc})
to be fixed by numerical simulations.

\begin{enumerate}
\item
{\it Jet head dynamics}. 
After the jet propagates inside the star,
it collides with the stellar envelope.
A reverse shock decelerates the jet, and
a forward shock runs into the stellar envelope.
The shocked region is called the jet head.
The jet head dynamics is determined by
the ram pressure balance between
the shocked jet and the shocked envelope,
both of which are given by the pre-shock quantities
through the shock jump conditions
\citep{Marti97,Meszaros01,Matzner03}:
\begin{equation}
h_j \rho_j c^2 \Gamma_{jh}^2 \beta_{jh}^2 + P_j 
= h_a \rho_a c^2 \Gamma_{h}^2 \beta_{h}^2 + P_a,
\end{equation}
where
$\Gamma_{jh}=\Gamma_j \Gamma_h (1-\beta_j\beta_h)$
is the relative Lorentz factor between the jet and the jet head
and $\beta_{jh}=(\beta_{j}-\beta_{h})/(1-\beta_{j}\beta_{h})$
is the corresponding relative velocity.
We can neglect the internal pressure of the jet $P_j$ 
for the strong reverse shock
and the pressure $P_a$ for the cold ambient matter.
Then, the jet head velocity is
\begin{equation}
\beta_h=\frac{\beta_j}{1+\tilde L^{-1/2}},
\end{equation}
where  
\begin{equation}
\tilde L \equiv \frac{h_j \rho_j \Gamma_j^2}{h_a \rho_a}
\simeq \frac{L_j}{\Sigma_j \rho_a c^3}
\label{eq:Ltilde}
\end{equation}
is the ratio of the energy density between the jet and the ambient medium.
In the last equality, we assume the cold ambient medium $h_a=1$
and use the jet cross-section $\Sigma_j = \pi r_j^2$ 
and the jet luminosity $L_j$.
For typical parameters of GRBs,
we have $\tilde L \ll 1$, i.e.,
a non-relativistic head velocity:
\begin{equation}
\beta_h \simeq \tilde L^{1/2}.
\label{eq:betah.nonrela}
\end{equation}
Hereafter, we consider the non-relativistic case,
that is appropriate for typical parameters.

\item
{\it Cocoon}. 
The shocked jet and the shocked envelope try to expand
and go sideways into a cocoon component.
The cocoon pressure is determined by
the injected energy divided by the cocoon volume \citep{Begelman89},
\begin{equation}
P_c=\frac{E}{3V_c}
=\frac{\eta}{3}
\frac{\int L_j (1-\beta_h) dt}{(\int \beta_h c\, dt)\, 
\pi (\int \beta_c c\, dt)^2},
\label{eq:pc}
\end{equation}
where $\eta$ is a parameter 
to correct the approximation of the cylindrical cocoon shape.
We use $\eta$ to absorb other approximations.
(For example, we represent the transverse velocity
by a single value, assume a spherical cocoon and
a power-law density profile in Equation~(\ref{eq_xi}), neglect
$z_*$ in Equation~(\ref{eq:zhatmax}), and so on.)
We determine $\eta$ by comparing the
analytical formulae with the numerical simulations.
Note that $1-\beta_h \approx 1$ for the non-relativistic head velocity.
The transverse velocity of the cocoon
is determined by the balance between the cocoon pressure 
and the ram pressure of the ambient medium:
\begin{equation}
\beta_c = \sqrt{\frac{P_c}{\bar \rho_a c^2}},
\label{eq:betac}
\end{equation}
where 
\begin{equation}
\bar \rho_a(z_h) = \frac{\int \rho_a dV}{V_c} \equiv \xi_a \rho_a(z_h),
\label{eq:xiadef}
\end{equation}
is the mean density of the medium.
Defining
\begin{equation}
\int \beta_h \, dt \equiv \xi_h \beta_h t \left(= \frac{z_h}{c}\right),
\quad
\int \beta_c \, dt \equiv \xi_c \beta_c t,
\label{eq:xihc}
\end{equation}
we can eliminate $\beta_c$ from Equations~(\ref{eq:pc}) and (\ref{eq:betac})
to obtain
\begin{equation}
P_c= \tilde L^{-1/4} \left(\frac{L_j \rho_a}{c t^2}\right)^{1/2}
\left(\frac{\eta \xi_a}{3\pi \xi_h \xi_c^2}\right)^{1/2}.
\label{eq:Pc2}
\end{equation}
If the density profile follows a power law $\rho_a \propto z^{-\alpha}$,
the coefficients, $\xi_a$, $\xi_h$ and $\xi_c$, become constant:
\begin{equation}
\label{eq_xi}
\xi_a=\frac{3}{3-\alpha}, \quad 
\xi_h=\xi_c=\frac{5-\alpha}{3}.
\label{eq:xialpha}
\end{equation}
Here we obtain $\xi_a$ in Equation~(\ref{eq:xiadef}) 
assuming a spherical cocoon with a radius $z_h$
for simplicity.
Although this is not a good approximation,
we adjust $\eta$ in Equation~(\ref{eq:pc})
to fit the analytic models with numerical results.
For $\xi_h$ and $\xi_c$,
we use Eqs.~(\ref{eq:betah2}), (\ref{eq:betac}) and (\ref{eq:Pc3}),
which yield
$\beta_h \propto t^{-2/5} (\beta_h t)^{\alpha/5} 
\propto t^{\frac{\alpha-2}{5-\alpha}}$
and $\beta_c \propto t^{-2/5} (\beta_h t)^{\alpha/5} \propto 
t^{\frac{\alpha-2}{5-\alpha}}$.

\item
{\it Collimation shock}.
If the ambient density is sufficiently high like in the stellar envelope,
the cocoon pressure becomes high enough to collimate the jet.
The initially expanding jet hits the first collimation shock
and its trajectory converges \citep{Bromberg11}.
For a constant cocoon pressure,
the conical jet becomes cylindrical after the collimation.
This process determines the cross-section of the jet
and thereby the jet Lorentz factor after the shock.
\end{enumerate}

The geometry of the collimation shock is determined by
the pressure balance between the jet and the cocoon
\citep{Komissarov97,Bromberg11},
\begin{equation}
h_{0} \rho_{0} c^2 \Gamma_{0}^2 \beta_{0}^2 \sin^2 \Psi + P_{0} = P_{c},
\label{eq:CSbalance}
\end{equation}
where the first term is the ram pressure of the jet 
normal to the shock surface
(see Figure~\ref{fig:jet_schematic}) and
the subscript $0$ stands for the unshocked jet.
Since the jet internal pressure decreases as $P_{0}\propto z^{-4}$
when the size grows, we neglect the term $P_{0}$.
For a small incident angle, we have a relation
\begin{equation}
\sin \Psi = \frac{r_s}{z} - \frac{d r_s}{dz} 
= z \frac{d}{dz}\left(\frac{r_s}{z}\right),
\label{eq:sinPsi}
\end{equation}
to the first order
(see Figure~\ref{fig:jet_schematic}),
where $r_s$ is the cylindrical radius of the shock position.
Then, Equation (\ref{eq:CSbalance}) 
gives a first-order ordinary differential equation.
Assuming that
$\beta_{0} \approx 1$ and
$L_j \simeq h_{0} \rho_{0} c^3 \Gamma_{0}^2 (\pi z^2 \theta_0^2)$,
we can integrate the geometry of the collimation shock as
\begin{equation}
r_s = \theta_0 z (1 + A z_*) - \theta_0 A z^2,
\label{eq:shockPconst}
\end{equation}
where $A$ is given by Equation~(\ref{eq:zhatmax}).
We note that we assume
a constant external pressure $P_c=$const.
For a decreasing external pressure, the shock geometry is different as shown
in Section~\ref{sec:model.postbreak}.

According to Equation~(\ref{eq:shockPconst}), 
the collimation shock 
expands to a maximum size at $\frac{dr_s}{dz}|_{z=z_{\max}}=0$,
and converges to $r_s(z=\hat z)=0$
where the maximally-expanding position $z_{\max}$
and the converging position $\hat z$ are given by
\begin{equation}
\hat z = 2 z_{\max}= A^{-1}+z_{*}
\simeq A^{-1} = \sqrt{\frac{L_{j} \beta_{0}}{\pi c P_c}}.
\label{eq:zhatmax}
\end{equation}
In the second to last equality
we assume that the collimation shock is initially small, $z_* \ll A^{-1}$.
As we can see from Figure~\ref{fig:jet_schematic},
the maximum size of the collimation shock
determines the cross-section of the jet after the collimation shock,
\begin{equation}
\Sigma_j(z>\hat z) \simeq \pi \left(\theta_0 z_{\max}\right)^2
\simeq \frac{L_j \theta_0^2}{4c P_c}.
\label{eq:Sigmaj}
\end{equation}
Combining with a general relation for a hot jet
(the radiation-dominated jet),
\begin{equation}
L_j \simeq 4 P_c \Gamma_{1}^2 \Sigma_j c,
\label{eq:hotjet}
\end{equation}
we obtain the Lorentz factor of the jet after the first collimation shock
\citep{Bromberg11},
\begin{equation}
\label{eq:gamma1}
\Gamma_{1} \simeq \frac{1}{\theta_0}.
\label{eq:gamma.const}
\end{equation}
After the collimation shock,
the jet is usually over-deflected to the axis,
resulting in an oblique shock inside the jet.
A high pressure region appears after the converging point
of the first collimation shock and then expands again.
The jet repeats a cycle of bounce, expansion, and collimation, 
like a spring under the pressure of the cocoon.
Because the supersonic jet is not synchronized with the cocoon,
oblique shocks arise in the jet.
In each collimation, the jet tries to expand
with the same angle $\sim 1/\Gamma_{0}$ 
and hence the Lorentz factor 
after a collimating oblique shock maintains its the same value.
The Lorentz factor after the $N$-th collimating oblique shock is
\begin{equation}
\Gamma_{N} \simeq \Gamma_{(N-1)} \simeq \dots \simeq \Gamma_{1} \simeq \frac{1}{\theta_0} \simeq \Gamma_{0},
\label{eq:gaj*}
\end{equation}
under  constant pressure inside a star.
The last equality is satisfied in our simulation 
because we inject a jet parallel to the $z$-axis
and the jet tries to expand with an angle $\theta_0 \sim 1/\Gamma_{0}$,
about the inverse of the initial Lorentz factor.

If the Lorentz factor were constant ($\Gamma \sim \Gamma_{0}$) 
even at the jet breakout,
the opening angle of the jet would be
the inverse of the Lorentz factor inside a star
(or the initial opening angle)
$\theta_{j} \sim \Gamma_{0}^{-1} \sim \theta_0$.
However this is not the case as discussed in Section
\ref{sec:model.postbreak}.

For comparison with the numerical results,
we express physical quantities by basic parameters, i.e.,
the jet luminosity $L_{j}$, the ambient density $\rho_a$,
the initial opening angle $\theta_0$, and time $t$.
First, we rewrite the head velocity and the cocoon pressure as
\begin{eqnarray}
\beta_h &\simeq & \tilde L^{1/2} \simeq
\left(\frac{L_j}{c^5 t^2 \rho_a \theta_0^4}\right)^{1/5}
\left(\frac{16}{3\pi}\frac{\eta \xi_a}{\xi_h \xi_c^2}\right)^{1/5},
\label{eq:betah2}
\\
P_c &=& \left(\frac{\rho_a^3 L_j^2 \theta_0^2}{t^4}\right)^{1/5}
\left(\frac{1}{6 \pi}\frac{\eta \xi_a}{\xi_h \xi_c^2}\right)^{2/5},
\label{eq:Pc3}
\end{eqnarray}
respectively, with Equations (\ref{eq:Ltilde}), (\ref{eq:betah.nonrela}), 
(\ref{eq:Pc2}) and (\ref{eq:Sigmaj}).
Then we can derive analytic formulae for
the jet head position with Equations.~(\ref{eq:xihc}) and (\ref{eq:betah2}),
the jet cylindrical radius with Equations.~(\ref{eq:Sigmaj}) and (\ref{eq:Pc2}),
and the converging point of the collimation shock
with Equations.~(\ref{eq:zhatmax}) and (\ref{eq:Pc2}) as
\begin{eqnarray}
z_h &=& \xi_h \beta_h c t
\simeq \left(\frac{t^3 L_j}{\rho_a \theta_0^4}\right)^{1/5}
\left(\frac{16}{3\pi}\frac{\eta \xi_a \xi_h^4}{\xi_c^2}\right)^{1/5}
\label{eq:zh}
\\
&\simeq&
1.4 \times 10^{10} \, {\rm cm}
\left(\frac{t}{1\, {\rm s}}\right)^{3/5}
\left(\frac{L_j}{10^{51}\, {\rm erg}\, {\rm s}^{-1}}\right)^{1/5}
\nonumber\\
 &\times& \left(\frac{\rho_a}{10^{3}\, {\rm g}\, {\rm cm}^{-3}}\right)^{-1/5}
\left(\frac{\theta_0}{0.1}\right)^{-4/5},
\label{eq:zh2}
\\
r_j &\equiv& \left(\frac{\Sigma_j}{\pi}\right)^{1/2} \simeq 
\left(\frac{t^4 L_j^3 \theta_0^8}{c^5 \rho_a^3}\right)^{1/10}
\left(\frac{3}{16\pi^{3/2}}\frac{\xi_h \xi_c^2}{\eta \xi_a}\right)^{1/5}
\nonumber\\
&\simeq&
2.4 \times 10^{8} \, {\rm cm}
\left(\frac{t}{1\, {\rm s}}\right)^{2/5}
\left(\frac{L_j}{10^{51}\, {\rm erg}\, {\rm s}^{-1}}\right)^{3/10}
\nonumber\\
&\times& \left(\frac{\rho_a}{10^{3}\, {\rm g}\, {\rm cm}^{-3}}\right)^{-3/10}
\left(\frac{\theta_0}{0.1}\right)^{4/5},
\label{eq:rj}
\\
\hat z &\simeq&
\left(\frac{t^4 L_j^3}{c^5 \rho_a^3 \theta_0^2}\right)^{1/10}
\left(\frac{6}{\pi^{3/2}}\frac{\xi_h \xi_c^2}{\eta \xi_a}\right)^{1/5}
\nonumber\\
&\simeq&
4.7 \times 10^{9} \, {\rm cm}
\left(\frac{t}{1\, {\rm s}}\right)^{2/5}
\left(\frac{L_j}{10^{51}\, {\rm erg}\, {\rm s}^{-1}}\right)^{3/10}
\nonumber\\
&\times& \left(\frac{\rho_a}{10^{3}\, {\rm g}\, {\rm cm}^{-3}}\right)^{-3/10}
\left(\frac{\theta_0}{0.1}\right)^{-1/5},
\label{eq:hatz}
\end{eqnarray}
respectively,
where we set $\alpha=2$ [$\xi_a=3$, $\xi_h=\xi_c=1$ in Equations~(\ref{eq:xialpha})]
and $\eta=0.01$ for the numerical values.
For a general density profile rather than a power-law form,
we solve the first line of Equation (\ref{eq:zh})
for the head position $z_h$,
where the quantities $\rho_{a}$, $\xi_a$, $\xi_h$ and $\xi_c$
in the right-hand side are also functions of $z_h$.
Here, for simplicity, we set $\xi_a$, $\xi_h$ and $\xi_c$
with Equation~(\ref{eq:xialpha})
using the density slope $\alpha=-d\ln \rho_a/d\ln z$
at $z=z_h$.

\begin{figure}
\epsscale{1.3}
\rotatebox{0}{\plotone{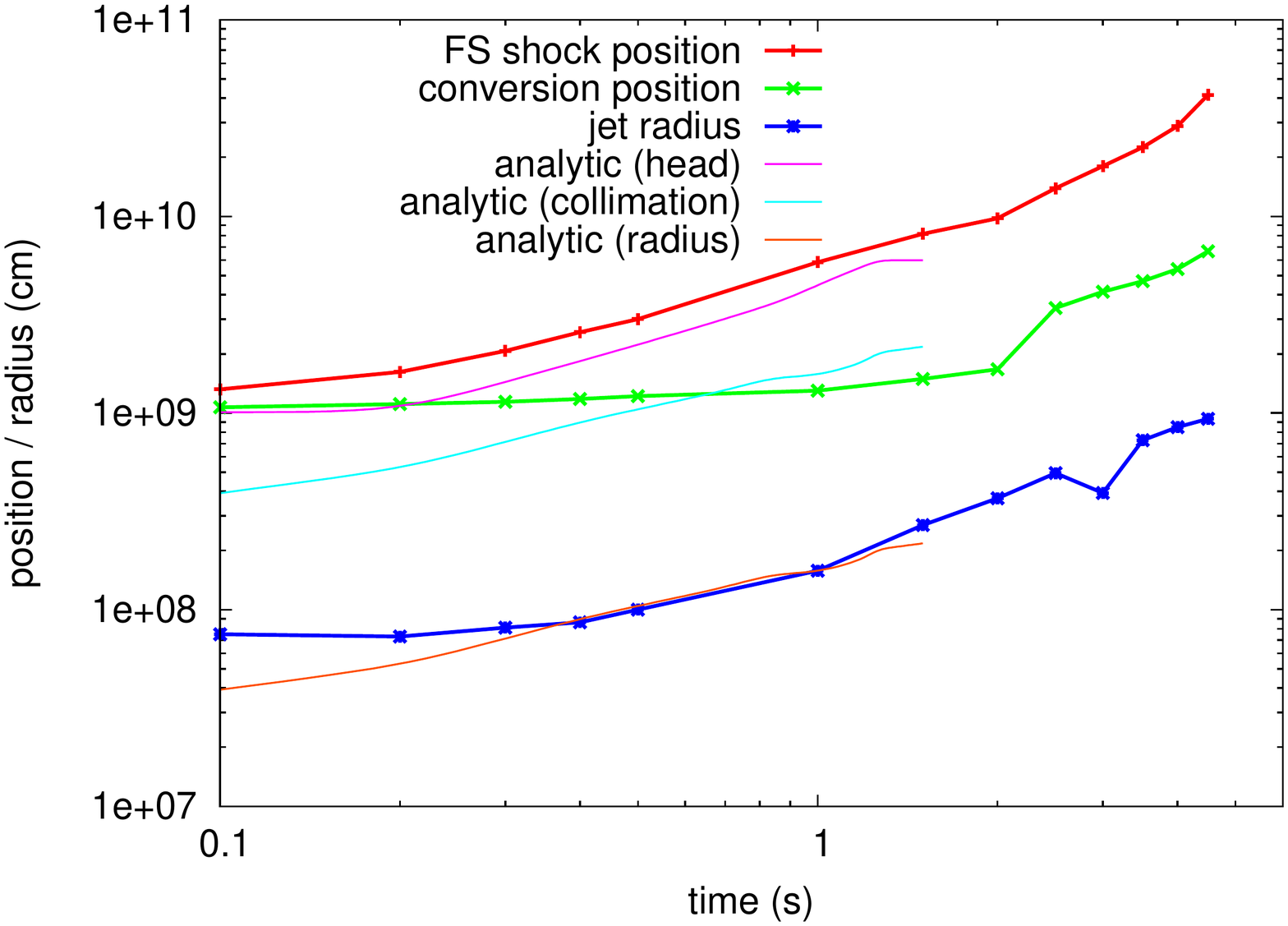}}
\caption{\label{fig:analytic}
Comparison of the numerical results 
with the analytical formulae for the jet propagation inside the star
(see Section~\ref{sec:model.prebreak}).
We show the forward shock position of the jet head $z_h$,
the converging position of the collimation shock $\hat z$,
and the maximum cylindrical radius of the jet 
$r_{s,\max} \simeq \theta_0 z_{\max}$.
Theoretical model lines are also plotted.
Since the slope of the radial mass density profile
starts to rapidly drop at $R\sim 6\times 10^{9}$~cm,
there is no analytical solution outside this point.
}
\end{figure}

As shown in Section~\ref{sec:results}, we can fit 
three analytic formulae in 
Equations~(\ref{eq:zh2})$-$(\ref{eq:hatz})
with the numerical calculations 
by adjusting one parameter, $\eta \sim 0.01$, in Equation~(\ref{eq:pc}).
Figure \ref{fig:analytic} shows
the comparison with analytic formulae and
the results of the hydrodynamic simulations,
i.e., the head position, the converging position,
and the jet radius,
for the model G5.0 ($\Gamma_{0}=5$ and
${\Delta z}_{\rm min}={\Delta r}_{\rm min}=10^7{\rm cm}$).
The numerical results are in good agreement with the analytical results.
The difference in the early evolution of the jet
is caused by the insufficient smallness of the initial jet size.
Since the power law index ($\alpha$) in the mass density profile
exceeds $\alpha > 3$ at a certain radius
and $\xi_a$ becomes infinity or negative in Equation~(\ref{eq_xi}),
we can not apply the analytical formula after $t=1.05$~s.
Figure~\ref{fig:analytic} is the confirmation
that 
the analytical physical picture of \citet{Bromberg11} is correct\footnote{
We introduce a parameter to fit the numerical results with the
analytic formula.
It is $\eta$ in Equation~(\ref{eq:pc}),
and turns out to be relatively
small ($\eta\sim 0.01$).
One of the reason for the smallness of $\eta$
is the weak dependence of the radii on $\eta$
in Equations.~(\ref{eq:zh})-(\ref{eq:hatz}).
These analytic
formulae for radii should have errors of a factor of $\sim 2$ because of the
reasons discussed below Equation~(\ref{eq:pc}). This factor corresponds to $1/2^5\sim 0.03$
for $\eta$,
which differs only by a factor of $\sim 3$ from the fitting result 
($\eta\sim 0.01$).
}
based on the numerical calculations 
that take great care with the baryon loading and the initial jet size.
Although the opening angle of the jet measured from the stellar origin
is small just before the breakout, the jet cannot maintain this
opening angle after the jet breakout because the jet is hot and would
try to expand with $\sim 1/\Gamma_0$ due to the thermal pressure.

\subsection{Jet Evolution in Decreasing External Pressure}
\label{sec:model.postbreak}

We numerically find that
the opening angle of the jet is not 
$\sim 1/\Gamma_{0}$ but $\sim 1/5\Gamma_{0}$
where $\Gamma_{0}$ is the Lorentz factor of the jet inside the star.
This is because the cocoon expands 
into outer space
and hence the cocoon pressure decreases outward 
after the jet breakout from the stellar surface.
In the decreasing pressure, the jet becomes less collimated
and hence more accelerated than that in the star.
Since the jet-breakout acceleration increases
the Lorentz factor $\Gamma$ before free expansion begins,
the jet opening angle determined by $\sim 1/\Gamma$
is narrower than the naive estimate.

The jet-breakout acceleration is broadly divided into two types,
i.e., with and without a shock.
Both types contribute equally to the Lorentz factor enhancement.
Since it is difficult to disentangle the two acceleration mechanisms,
we discuss these cases separately below, assuming each mechanism is
dominant.

First, we consider 
the jet-breakout acceleration
by evaluating the Lorentz factor after the 
collimating oblique shock.
The cocoon pressure that is decreasing outward
is expressed by
\begin{equation}
P_c(z)=P_{*}\left(\frac{z}{z_{*}}\right)^{-\lambda}.
\end{equation}
Under this pressure,
we can solve the geometry of the collimation shock,
as in Section \ref{sec:model.prebreak}
with Equations~(\ref{eq:CSbalance}) and (\ref{eq:sinPsi}):
\begin{equation}
r_s = \theta_0 z \left(1 + \frac{A_{*} z_{*}}{1-\frac{\lambda}{2}} \right) 
- \frac{\theta_0 A_{*} z^2}{1-\frac{\lambda}{2}}
\left(\frac{z}{z_{*}}\right)^{-\frac{\lambda}{2}},
\end{equation}
where $z_{*}$ is the initial position of the collimation shock and
\begin{equation}
A_{*}=\sqrt{\frac{\pi c P_{*}}{L_j \beta_{0}}}.
\end{equation}
We show the shock geometry in Figure~\ref{fig:jet_schematic}.
Note that the shock geometry is not a parabola,
in contrast with the constant pressure case in Equation~(\ref{eq:shockPconst}).
The collimation shock 
expands to a maximum cylindrical radius 
at $\frac{dr_s}{dz}|_{z=z_{\max}}=0$,
where the maximally-expanding position is given by
\begin{equation}
\frac{z_{\max}}{z_*}= \left[
\left(\frac{2-\lambda}{A_{*} z_{*}}+2\right) 
\frac{1}{4-\lambda}\right]^{\frac{2}{2-\lambda}}.
\end{equation}
Here $\lambda < 2$ is necessary for the shock to converge.
At the position $z_{\max}$,
the general relation for a hot jet in Equation~(\ref{eq:hotjet})
is given by
\begin{eqnarray}
L_j 
\simeq 
4 \left[P_{*} \left(\frac{z_{\max}}{z_{*}}\right)^{-\lambda}\right]
\Gamma_{1}^2 
\left(\pi \theta_0^2 z_{\max}^2\right) c.
\end{eqnarray}
This yields the Lorentz factor after the collimation shock,
\begin{eqnarray}
\Gamma_{1} 
\simeq \frac{1}{\theta_0}
\times {\cal A},
\label{eq:Gamma1}
\end{eqnarray}
which is larger than that for a constant pressure case 
in Equation~(\ref{eq:gamma.const}) by
\begin{equation}
{\cal A} = \frac{4-\lambda}{4-2 \lambda + 4 A_{*} z_{*}}
\simeq \frac{4-\lambda}{4-2 \lambda}
\sim 5,
\quad {\rm if}\quad 
\lambda \sim 1.8,
\end{equation}
for small $A_{*} z_*$.
To be precise, $\theta_0$ is not
the initial opening angle of the jet here,
but the opening angle of the jet expanding
to the last collimating oblique shock,
which is the inverse of the Lorentz factor inside a jet 
$\sim \Gamma_{0}^{-1}$ and thereby turns out to be $\theta_0$
from Equations~(\ref{eq:gamma.const}) and (\ref{eq:gaj*}).

The numerical calculations show that
the total acceleration factor is 
${\cal A} \sim 5$ (see Figure~\ref{fig:analytic}),
about half of which is achieved at the collimation shock
and the other half of which is obtained later.
A factor $\sim 2.5$ can be explained by $\lambda \sim 1.5$.
Although it is difficult to estimate the exact acceleration factor 
${\cal A}$ analytically,
we can see that ${\cal A}$ is mainly determined by
the slope $\lambda$ of the external pressure profile.
In our case, the external pressure profile is shaped by
the cocoon expansion to outer space,
which does not depend on the jet properties so much.
Therefore it is natural that similar acceleration factors
are obtained for the different initial conditions
in our numerical calculations.

Next, we consider a jet-breakout acceleration 
without shocks 
(i.e., an adiabatic jet)
as the other extreme.
The jet expands in a decreasing pressure,
decreasing its temperature as radiation, i.e.,
$T' \propto V'^{-1/3}$,
where $V'$ is the comoving volume of the jet.
If the pressure in the jet balances with the pressure in the cocoon,
i.e.,
$T'^4 \propto P_c$,
the jet does not perform work
and hence the energy is conserved and $\Gamma T'^4 V' \sim$ const.
Therefore the jet accelerates as the size grows
according to \citep{KI+11}
\begin{equation}
\Gamma \propto T'^{-1} \propto P_c^{-1/4} \propto z^{\lambda/4}.
\label{eq:gammaz}
\end{equation}
In the numerical calculations, we identify
such an adiabatic evolution 
after the collimation shock, in particular at the periphery of the jet.
For $2 < \lambda < 4$, the cocoon pressure does not causally affect
the jet interior \citep{KI+11}
but can still affect the periphery of the jet.
Note that the off-center origin makes the pressure profile
(Figure~\ref{fig:1Dpressure}(b))
shallower
than that shown in Figure~\ref{fig:1Dpressure}(a),
which the assumes a stellar center origin.

In order to estimate the acceleration factor of the Lorentz factor,
we need to know how much the jet expands before entering a free expansion phase.
In this regard, we should note that the jet expands from the breakout position, 
that is, the fireball of the jet 
is off-centered by the cocoon confinement.
This means that the fireball size should not be
measured from the center of the star.
Instead, the effective center of the fireball
is located at a distance $\sim \Gamma_{0} r_{j} \sim r_{j}/\theta_0$ 
inward from the stellar surface (breakout point)
because the jet tries to expand with an opening angle of $1/\Gamma_{0}$
and the expanding surface has an initial cylindrical radius of 
$r_{j}$ in Equation~(\ref{eq:rj}).
So, the initial fireball size is 
\begin{eqnarray}
r_{0} \sim \Gamma_{0} r_{j} \sim r_{j}/\theta_0.
\end{eqnarray}
If the fireball expands to a size of the stellar radius
$R \sim 4 \times 10^{10}$ cm,
the expansion factor is about $\sim 10$ times
for the typical parameters in Equation.~(\ref{eq:rj})
and hence the Lorentz factor grows by a factor
$\sim 3$ for $\lambda \sim 2$,
according to Equation~(\ref{eq:gammaz}).
Thus, the adiabatic expansion can explain
a part of the jet-breakout acceleration observed 
in the numerical calculations.
Here, a parameter dependence of the expansion factor is weak
since the initial fireball size is
$r_0 \propto L_j^{1/6} \rho_a^{-1/6} \theta_0^{1/3}$
at the breakout time (when $R \sim z_h$)
with Equations~(\ref{eq:zh2}) and (\ref{eq:rj}).
Note that the jet-breakout acceleration looks very rapid
at first glance
if we do not note the off-center effect
(i.e., the radius measured from the stellar center is only doubled).

\section{DISCUSSION}
\label{sec:discussion}
\subsection{Lorentz Factor of the Jet in a Star}
\begin{figure}
\epsscale{1.2}
\plotone{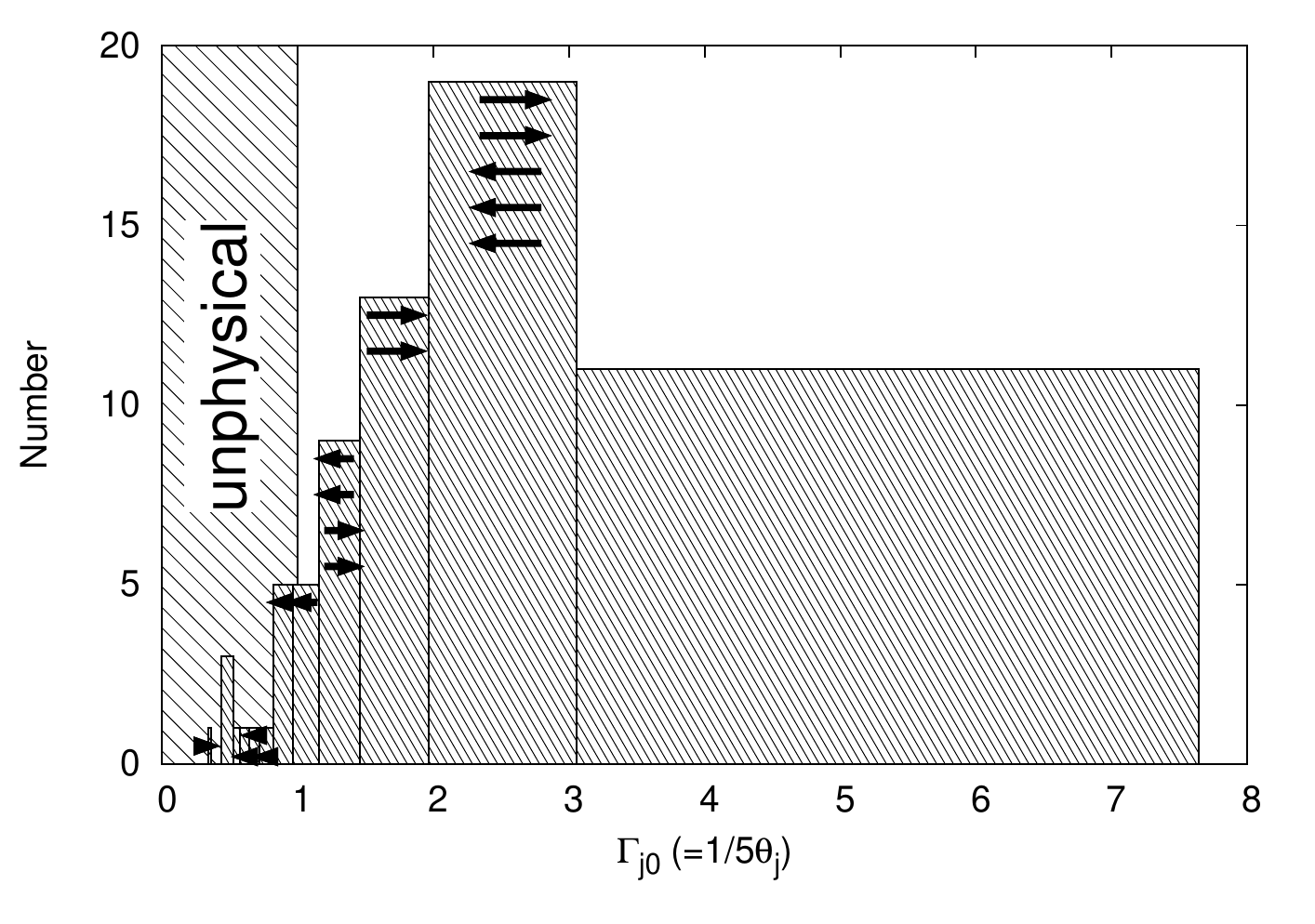}
\caption{\label{fig:fong12}
Distribution of the initial Lorentz factor of the jet
inferred by the observations 
and the relation $\Gamma_{0}=1/5\theta_j$
in Equation~(\ref{eq:thj}).
Observational data ($\theta_j, N$)
are taken from Figure~6 in \citet{Fong12}.
Since the Lorentz factor should be greater than unity,
the region with $\Gamma_{0}<1$ is shaded
as an unphysical zone.
}
\end{figure}

The opening angle of the GRB jet is usually measured by
observing a jet break in the afterglow lightcurve
\citep{Racusin09,Racusin11,Fong12}.
Our numerical calculations show that
a jet opening angle is related to
the Lorentz factor inside a star by
\begin{equation}
\theta_{j} \sim \frac{1}{5\Gamma_{0}}.
\label{eq:thjga0}
\end{equation}
By applying this formula, 
we can infer the Lorentz factor inside a star
(or the initial opening angle)
from the observed opening angle of the GRB jet.
Figure \ref{fig:fong12} shows the estimated Lorentz factor inside a star.
The result suggests that the jet is mildly relativistic in a star
(or the initial opening angle is O(0.3-0.5) rad).

A slow jet inside a star is a preferable condition
for the survival of nuclei in the jet,
which may explain the observed ultra-high energy cosmic ray nuclei
\citep{Murase08,Wang08,Metzger11,Horiuchi12}.

Equation (\ref{eq:thjga0}) also implies that 
the maximum opening angle is obtained by setting $\Gamma_0 \sim 1$ as
\begin{equation}
\theta_{j,\max} \sim 1/5=0.2 \sim 12^{\circ},
\label{eq:thjmax}
\end{equation}
if the jet is radiation dominated at the breakout.
However several events violate this maximum, as shown in Figure~\ref{fig:fong12}
\citep{Racusin09,Racusin11,Fong12}.
There are two
possibilities to this problem.
The first is to consider 
a baryon-rich slow sheath surrounding a central jet.
A baryon-rich flow cannot accelerate to a large Lorentz factor,
and if the Lorentz factor of the baryon-rich sheath is less than $\sim 5$,
the opening angle of the sheath can be larger than $\sim 0.2$.
Note that the central jet should have $h\Gamma > 100$ 
to avoid the compactness problem.
\footnote{A potential third possibility is that the outflow is initially
non-relativistic and expands to a relativistic speed. The
non-relativistic flow expands to an angle larger than the relativistic
case, $\theta_0 \sim \pi  > 1/\Gamma_0$,
so that the final opening angle after
the breakout might be also larger than the maximum value in Equation~(\ref{eq:thjmax}).
However, the collimation of the spherically expanding flow cannot be
treated by the formulae in this paper and \citet{Bromberg11}.
So, we leave this possibility for the future studies.
}
The second is to consider long-acting engine activity (e.g.,
\citet{Ioka05}).
After several
tens of seconds in Equation~(\ref{eq:duration}), the jet is no longer confined by the
cocoon and the opening angle can widen to $\theta_j\sim 1/\Gamma_0$ without
the factor of $\sim 5$ (see the next section).
If the jet energy is dominated by the wide opening angle component,
the wide component determines the opening angle of the jet obtained from
the afterglow observations.

\subsection{Origin of the GRB Duration}
\label{sec:duration}
In our new picture with the jet-breakout acceleration
in Figure~\ref{fig:break},
the GRB duration would be determined by
the sound crossing time of the cocoon,
which is about $t_{sc}\sim R_*/c\beta_c$,
where 
the sound velocity in the cocoon is about $c \beta_c$ in Equation~(\ref{eq:betac}).
At $t<t_{sc}$,
the cocoon persists around the star and provides pressure for
collimating the jet into an opening angle
$\theta_j \sim 1/5\Gamma_0$ in
Equation~(\ref{eq:thj}).
However, at $t>t_{sc}$,
the cocoon pressure decreases
and thereby the jet is no longer confined.
The opening angle of the
jet increases to $\theta_j \sim 1/\Gamma_0$,
which is determined by the free expansion
without the jet-breakout acceleration
(see also Morsony et al. 2007).
Then, the apparent luminosity of the GRB jet is
reduced by a factor of $\sim 5^2 \sim 25$.
Even if the jet injection continues after $t>t_{sc}$,
we observe that the GRB terminates.
Therefore we expect the observed duration to be
\begin{eqnarray}
T_{90} \sim t_{sc} \equiv R_*/c\beta_c.
\label{eq:duration}
\end{eqnarray}
With Equations(\ref{eq:betac}), (\ref{eq:xiadef}),
and (\ref{eq:Pc3}), we have
\begin{eqnarray}
T_{90} \sim 
R_* \left( \rho_a t^2\over L_j \theta_0 \right)^{1/5}
\left( (6\pi)^2 \xi_a^3 \xi_h^2 \xi_c^4\over \eta^2 \right)^{1/10}.
\end{eqnarray} 
Here it is appropriate to set the time $t$ to the breakout time
determined by $z_h=R_*$ in Equation~(\ref{eq:zh}),
which is
\begin{eqnarray}
t& = &t_{br} \equiv 
\left(R_*^5 \rho_a \theta_0^4 \over L_j\right)^{1/3}
\left( {3\pi\over 16}  
{\xi_c^2 \over \eta \xi_a \xi_h^4 }\right)^{1/3}\nonumber \\
& \sim & 0.58 \, {\rm s}
\left(\frac{R_*}{10^{10}\, {\rm cm}}\right)^{5/3}
\left(\frac{L_j}{10^{51}\, {\rm erg}\, {\rm s}^{-1}}\right)^{-1/3}
\nonumber\\
&\times& \left(\frac{\rho_a}{10^{3}\, {\rm g}\, {\rm cm}^{-3}}\right)^{1/3}
\left(\frac{\theta_0}{0.1}\right)^{4/3}.
\label{eq:br}
\end{eqnarray}
Then we have
\begin{eqnarray}
T_{90} & = &\left( {R_*^5 \rho_a \theta_0 \over L_j}\right)^{1/3}
\left( {(3\pi)^2\over 4}  {\xi_a \xi_c^4 \over \xi_h^2
 \eta^2}\right)^{1/6}\nonumber \\
& \sim & 20\, {\rm s}
\left(\frac{R_*}{10^{10}\, {\rm cm}}\right)^{5/3}
\left(\frac{L_j}{10^{51}\, {\rm erg}\, {\rm s}^{-1}}\right)^{-1/3}
\nonumber\\
&\times& \left(\frac{\rho_a}{10^{3}\, {\rm g}\, {\rm cm}^{-3}}\right)^{1/3}
\left(\frac{\theta_0}{0.1}\right)^{1/3},
\label{Eq:T90}
\end{eqnarray}
which is consistent with the observed GRB duration $T_{90} \sim 10$ sec.
We again 
set $\alpha=2$ [$\xi_a=3$, $\xi_h=\xi_c=1$ in Equation~(\ref{eq:xialpha})]
and $\eta=0.01$ for the numerical values.

\subsection{Amati and Yonetoku Relations}

Based on our numerical and analytical modeling,
we can evaluate the initial condition of a jet
that is just expanding freely
and infer the observational characteristics of the jet
based on the photospheric model
\cite[e.g.,][]{Peer07,KI+07}.
First, since we are now able to estimate 
the opening angle of the jet in Equation~(\ref{eq:thjga0}),
we can assess the isotropic luminosity of the jet:
\begin{equation}
L_{\rm iso} = \frac{4 L_{j}}{\theta_{j}^2}
\propto L_{j} \Gamma_{0}^2.
\label{eq:Liso}
\end{equation}
We can also obtain the observed temperature of a jet
if it is radiation dominated:
\begin{equation}
E_{\rm peak} \sim T_{\rm obs} \propto \Gamma_{0} T' \propto \Gamma_{0} P_c^{1/4},
\label{eq:Epeak}
\end{equation}
which may be identified with the spectral peak energy $E_{\rm peak}$
of a GRB in the photosphere model.
In the actual observations of GRBs,
there is a relation between
the isotropic luminosity and the observed temperature,
\begin{equation}
E_{\rm peak} \propto L_{\rm iso}^{1/2} \quad,
\end{equation}
the so-called Yonetoku relation \citep{Yonetoku04}.

Let us show that the Yonetoku relation
can be reproduced if we assume that the total jet
luminosity is propositional to the Lorentz factor inside the star
\begin{equation}
L_{j} \propto \Gamma_{0},
\label{eq:constMdot}
\end{equation}
that is, the mass outflow rate is $\dot{M} = L_j/\Gamma_{0} \sim $ const.
First, the above equation gives $L_{\rm iso} \propto \Gamma_{0}^{3}$
with Equation~(\ref{eq:Liso}).
Next, substituting Equations~(\ref{eq:Liso}) and (\ref{eq:Epeak}) 
into Equation~(\ref{eq:hotjet}) 
(where we should read $\Gamma_1$ as $\Gamma_0$)
yields
\begin{equation}
E_{\rm peak} \propto L_{\rm iso}^{1/4} \Sigma_{j}^{-1/4}.
\end{equation}
The jet breakout occurs when $z_h \sim R_*$
at the time
\begin{eqnarray}
t \propto L_j^{-1/3} \theta_0^{4/3},
\end{eqnarray}
obtained from Equation~(\ref{eq:zh2}) (see also Equation~(\ref{eq:br})).
At this time the jet cross-section follows
\begin{eqnarray}
\Sigma_{j} \propto t^{4/5} L_j^{3/5} \theta_0^{8/5}
\propto L_{j}^{1/3}\theta_0^{8/3},
\end{eqnarray}
from Equation~(\ref{eq:rj})\footnote{
We assume that the structure of the progenitor star is similar
and neglect the dependence on it.}
.
Then, noting $\theta_0 \sim \Gamma_{0}^{-1}$, we have
\begin{equation}
E_{\rm peak} \propto L_{\rm iso}^{1/4} \Gamma_{0}^{7/12}
\propto L_{\rm iso}^{4/9},
\end{equation}
which is close to the Yonetoku relation.

In addition, we may be also able to reproduce the Amati relation
\citep{Amati02},
\begin{equation}
E_{\rm peak} \propto E_{\rm iso}^{1/2} \quad (\mbox{Amati\ relation}).
\end{equation}
We can think that the GRB duration is
roughly given by the sound-crossing time 
across the cocoon of the stellar size in Equation~(\ref{Eq:T90}),
\begin{eqnarray}
T_{90} \sim \frac{R_{*}}{c \beta_{c}}
\propto t^{2/5} L_{j}^{-1/5} \theta_{0}^{-1/5}
\propto L_{j}^{-1/3} \theta_{0}^{1/3}
\propto L_{\rm iso}^{-2/9},
 \end{eqnarray}
with Equations~(\ref{eq:betac}) and (\ref{eq:Pc3}),
because the cocoon pressure decreases after this time,
leading to less confinement,
a larger opening angle, and a smaller isotropic luminosity of the jet.
Note that the weak correlation between $T_{90}$ and $L_{\rm iso}$
is actually observed.
Then, we can estimate the isotropic energy as
\begin{equation}
E_{\rm peak} \propto L_{\rm iso}^{4/9}
\propto (L_{\rm iso} T_{90})^{4/7} \sim E_{\rm iso}^{4/7},
\end{equation}
which is also similar to the Amati relation.
This is killing two birds with one stone,
that is,
we explain the slopes of two relations (the Amati and Yonetoku relations)
by only one assumption in Equation~(\ref{eq:constMdot}).
We leave the explanation of the normalization factor for future studies.

\citet{Lazzati13} reproduced the Amati relation in the context of
the photosphere model by combining their numerical results with an
analytical model for estimating the peak energy.
It is not easy to compare the \citet{Lazzati13} model with our analytic model.
Since in \citet{Lazzati13},
most radiation is released in the coasting phase,
in which the most thermal energy has been converted to kinetic energy,
the temperature estimation is different from our estimates.
The second is that
the viewing angle dependence mainly
produces the Amati relation in \citet{Lazzati13},
which is different from our cases.

\section{SUMMARY}
\label{sec:summary}

In this paper we have explored the dynamics of GRB jets
from collapsars by performing
two-dimensional relativistic hydrodynamic simulations
as well as analytical modeling.
We have followed the jet propagation all the way
from the progenitor star
through the jet breakout to the free expansion,
implementing probe particles to 
trace the Lagrangian motion of the fluid elements.
This enables us to connect the final jet appearance 
to the initial jet conditions near the central engine.

We have found that the jet opening angle after the jet breakout
is about $\theta_j \sim 1/5\Gamma_0$ in Equation~(\ref{eq:thj})
and Figure~\ref{fig:angle.gamma},
where $\Gamma_0$ is the initial Lorentz factor of the jet
injected into the progenitor star.
This value is smaller than the naive expectation of $\theta_j \sim 1/\Gamma_0$
in Figure~\ref{fig:break},
where we thought that
the opening angle was determined by the inverse of the Lorentz factor
and that the Lorentz factor maintains its the initial value of $\Gamma_0$
for a cylindrical jet.
Actually, this is partly correct.
The jet becomes cylindrical under the nearly homogeneous pressure of the cocoon
after crossing the first collimation shock.
The Lorentz factor after the collimation shock
is $\sim \Gamma_0$ and largely stays constant before the jet breakout,
according to our simulations.
However, we have identified the ``jet-breakout acceleration''
just before and after the jet breakout.
This occurs because the pressure profile of the cocoon can not remain
constant but decreases outward as the cocoon expands to outer space.
The cocoon still confines the jet to some extent
near the stellar radius, while
the jet expands its cylindrical radius with
increasing its Lorentz factor by a factor $\sim 5$ before a free expansion.
Therefore the jet opening angle becomes narrow,
which is determined by the inverse of 
the Lorentz factor at the base of the free expansion,
as explicitly shown by the numerical simulations.
The opening angles are roughly constant over time
with a factor $\sim 2$ fluctuation in Figure~\ref{fig:angle.time}.

We have also analytically modeled
the jet-breakout acceleration.
The jet-breakout acceleration occurs with and without 
the collimating oblique shock,
and both are equally important.
For the former case, we solve the structure of the collimating oblique shock
in a decreasing pressure profile
and obtain the Lorentz factor after the shock
in Equation~(\ref{eq:Gamma1}).
The post-shock Lorentz factor is enhanced appreciably 
for a pressure slope close to
$\lambda \sim 2$ in $P \propto z^{-\lambda}$.
The latter case happens after the last collimating oblique shock,
even for $\lambda > 2$ near the periphery of the jet
in Equation~(\ref{eq:gammaz}).

We have also compared our numerical results with 
the analytical formulae for jet propagation inside the star 
presented by \citet{Bromberg11},
and have confirmed a good agreement.
For later use, we have calibrated 
the model parameter with the numerical results.
We can now calculate the jet evolution relatively precisely with ease,
such as the jet head position, the jet cylindrical radius,
and the converging position of the collimation shock,
for a wide range of initial conditions.

We have paid special attention to the numerical diffusion
of the baryon loading into the jet through the discontinuity,
which can entirely change the jet propagation.
We have also taken the initial cylindrical radius of the jet
to be sufficiently smaller than the radius after the first collimation shock,
because a large initial radius slows down the jet propagation.
For these purposes, we have performed 
one of the highest resolution calculations so far.

The post-breakout jet shows a hollow-cone angular structure.
The edge is relatively sharp with an exponential drop.
The bright rim is produced by the shock between the expanding jet and
the high pressure cocoon before the free expansion.

To understand the jet evolution, it is important to note that
the jet expands off-center as a result of the cocoon confinement.
If the expansion origin is the stellar center,
the fireball would feel a steep pressure profile of the cocoon pressure
and the acceleration would be slow.
In addition, the opening angle should be measured
from the off-center origin for precise analyses.

We have also applied our results to the observations.
First we infer the initial Lorentz factor $\Gamma_0$ of the jet
injected at the central engine
by using the observed opening angles in Figure~\ref{fig:fong12}.
The distribution of $\Gamma_0$ peaks at around $\sim 2$--$3$.
Second, our result suggests the existence of
a maximum opening angle for a high-entropy jet,
$\theta_{j,\max} \sim 1/5 \sim 12^{\circ}$,
in Equation~(\ref{eq:thjmax}).
However several bursts violate this maximum value.
This may imply a two-component jet with a baryon-rich slow sheath
surrounding a central jet,
or a long-lasting jet after the GRB prompt
emission.

The opening angle evolution with the jet-breakout
acceleration is also important for determining
the observed duration of GRB.
In particular, the GRB duration is given by
the sound crossing time of the cocoon in Equation~(\ref{eq:Pc3}).
Before this time, the cocoon continues to
exist around the jet and confines it into an opening angle
$\sim 1/5\Gamma_0$, while after that, the jet expands freely and the opening
angle increases to $\sim 1/\Gamma_0$.
This reduces the apparent luminosity of the GRB,
effectively terminating the observed GRB.

We have also derived the slopes of the Amati and Yonetoku spectral relations
by applying our results to 
the photosphere of the jet that is expanding freely after the jet breakout.
We explain the slopes of both the relations with only one assumption
that the jet luminosity is proportional to the initial
Lorentz factor, $L_j \propto \Gamma_0$, in Equation~(\ref{eq:constMdot}),
i.e., the mass outflow rate is independent of the jet luminosity,
$\dot M=L_j/\Gamma_0 \sim$ const.
The fireball temperature becomes different from 
the value at the central engine
after the jet propagation through the star.
Thus the confinement by the cocoon and the off-center expansion
of the jet
may be the missing pieces for the photosphere model so far.

In the future, it will be interesting to study 
the long-term evolution of the jet studied in this paper
and 
the evolution of the low-luminosity jet and the two-component jet.
It will  also be important to investigate
jet propagation in a huge progenitor
such as the population III GRBs 
\citep{SI11,Nagakura12}
and ultra-long GRBs
\citep{Levan13,Murase13}.
It will be interesting to perform magnetohydordynamic simulations to
study the effect of magnetic fields on the jet dynamics and the
opening angle of the jet. For example, \citet{Mignone10} have
performed magnetohydordynamic simulations in the context of active
galactic nucleus jets.
Three-dimensional numerical hydrodynamic
simulation including precession or other effects
will also be interesting (for example, see \citet{Lopez13}).

\

\acknowledgments
We thank A. Heger for his kindness to allow us to use
his progenitor models.
We thank the anonymous referee for constructive comments
that improved our manuscript.
This work
is supported by a Grant-in-Aid for Scientific Research from
the Ministry of Education, 
Culture, Sports, Science and Technology (MEXT) of Japan
and athe Japan Society for the Promotion of Science(JSPS)
(20105005 (AM), 24103006, 24000004, 22244030, 21684014 (KI)).
Numerical simulations were carried out on SR-16000, at YITP, Kyoto University,
on the Space Science Simulator (NEC SX9) at JAXA, 
and on the XT4 System at CFCA at NAOJ.
This work was supported in part by the Center for the Promotion of
Integrated Sciences (CPIS) of Sokendai. The page charge of this paper is partly supported by CFCA at NAOJ.


\end{document}